\documentclass[sort&compress]{elsarticle}

\usepackage{fullpage}
\usepackage{multirow}
\usepackage{lineno}
\usepackage{enumerate}
\usepackage{tikz}
\usepackage{url}
\usepackage{stmaryrd}

\usepackage{nomencl}
\usepackage{amsmath,mathtools}
\usepackage{amsfonts,bm}
\usepackage{multirow}
\usepackage{subfig}

\usepackage{graphicx}
\usepackage{booktabs}
\usepackage{makecell}
\usepackage{siunitx}
\usepackage[skip=0.6\baselineskip]{caption}

\usepackage{xpatch}
\let\oldequation\equation
\let\oldendequation\endequation
\renewenvironment{equation}
  {\linenomathNonumbers\oldequation}
  {\oldendequation\endlinenomath}
  
\renewcommand\nomgroup[1]{%
  \item[\itshape%\bfseries
  \ifstrequal{#1}{A}{Symbols}{%
  \ifstrequal{#1}{B}{Roman Letters}{%
  \ifstrequal{#1}{C}{Greek Letters}{%
  \ifstrequal{#1}{D}{Abbreviations}{}}}}%
]}

\usepackage{floatrow}
\usepackage[commentmarkup=footnote]{changes} 
\usepackage{xcolor}
\usepackage{bm}
\usepackage{url}

\newcommand{\region}[1]{\langle #1 \rangle}

\newcommand{\bbR}{\mathbb {R}}
\newcommand{\bx}{\bm{x}}
\newcommand{\by}{\Delta \bm{x}}
\newcommand{\bz}{\bm{z}}
\newcommand{\bu}{\mathbf{u}}
\newcommand{\bq}{\mathbf{q}}
\newcommand{\bw}{\bm{c}}
\newcommand{\cO}{\mathcal{R}}
\newcommand{\cR}{\mathcal{Q}}
\newcommand{\cQ}{\mathcal{Q}}
\newcommand{\cG}{\mathcal{G}}
\newcommand{\cX}{\mathcal{X}}
\newcommand{\cU}{\mathcal{U}}
\newcommand{\cW}{\mathcal{C}}
\newcommand{\cC}{\mathcal{C}}
\newcommand{\cD}{\mathcal{D}}
\newcommand{\cL}{\mathcal{L}}
\newcommand{\vnorm}[1]{|#1|}
\newcommand{\tnorm}[1]{\|#1\|}
\newcommand{\vstack}[1]{\llbracket #1\rrbracket}

\graphicspath{{./figs/}}

\usepackage{array}
\newcolumntype{P}[1]{>{\centering\arraybackslash}m{#1}}

\begin{document}
\begin{frontmatter}
  
\title{Frame-independent vector-cloud neural network for nonlocal constitutive modeling on arbitrary grids}

  \author[vt]{Xu-Hui Zhou\fnref{cofirst}}
  %\ead{xuhuizhou@vt.edu}
    
  \author[pu]{Jiequn Han\fnref{cofirst}\corref{cor}}
  \ead{jiequnhan@gmail.com}
  \cortext[cor]{Corresponding author}

  \author[vt]{Heng Xiao}
  \ead{hengxiao@vt.edu}
  
  \fntext[cofirst]{Contributed equally}
  
  \address[vt]{Kevin T. Crofton Department of Aerospace and Ocean Engineering, Virginia Tech, Blacksburg, VA 24060, USA}
  \address[pu]{Department of Mathematics, Princeton University, Princeton, NJ 08544, USA}

 \begin{abstract}
    
    Constitutive models are widely used for modeling complex systems in science and engineering, where first-principle-based, well-resolved simulations are often prohibitively expensive. For example, in fluid dynamics, constitutive models are required to describe nonlocal, unresolved physics such as turbulence and laminar-turbulent transition. However, traditional constitutive models based on partial differential equations (PDEs) often lack robustness and are too rigid to accommodate diverse calibration datasets. We propose a frame-independent, nonlocal constitutive model based on a vector-cloud neural network that can be learned with data. The model predicts the closure variable at a point based on the flow information in its neighborhood. Such nonlocal information is represented by a group of points, each having a feature vector attached to it, and thus the input is referred to as vector cloud. The cloud is mapped to the closure variable through a frame-independent neural network, invariant both to coordinate translation and rotation and to the ordering of points in the cloud. As such, the network can deal with any number of arbitrarily arranged grid points and thus is suitable for unstructured meshes in fluid simulations. The merits of the proposed network are demonstrated for scalar transport PDEs on a family of parameterized periodic hill geometries. The vector-cloud neural network is a promising tool not only as nonlocal constitutive models and but also as general surrogate models for PDEs on irregular domains.
\end{abstract}
 
  
\end{frontmatter}

%\linenumbers

\section{Introduction}
\label{sec:intro}

Constitutive models are widely encountered in science and engineering when the computational costs of simulating complex systems are often prohibitively high. Taking industrial computational fluid dynamics (CFD) for example, constitutive models such as Reynolds stress models and eddy viscosity models are often used to describe unresolved turbulence and to \emph{close} the Reynolds averaged Navier--Stokes (RANS) equations for the mean flows fields, which are of ultimate interest in engineering design. As such, they are also referred to as ``closure models'', a term that is used interchangeably in this work. Closure models are also used to describe laminar-turbulent transitions, which typically occur near the leading edge of airfoils or gas engine turbines. Despite the enormous growth of available computational resources in the past decades, even to this day such closure models are still the backbone of engineering CFD solvers. However, these models are typically based on a hybrid of partial differential equations (PDEs) and algebraic relations, which are difficult to calibrate. In particular, it is challenging to accommodate diverse sets of calibration data, especially those from realistic, complex flows. 

In the past few years, the emergence of neural networks and other machine learning models has led to unprecedented opportunities for constitutive modeling. Among the most attractive features of neural-network-based models is their expressive power, which could enable flexible constitutive models to be calibrated on a wide range of datasets and consequently lead to a \textit{de facto} universal model, or at least a \emph{unified} model with automatic internal switching depending on the regime. This has long been an objective of constitutive model development such as in turbulence modeling. In light of such strengths and promises, researchers  have made efforts to develop data-driven, machine-learning-based constitutive models in a wide range of application areas such as turbulence modeling~\cite{singh16machine,ling16reynolds,wang17physics-informed,wu2018physics-informed,yang2019predictive,schmelzer2020discovery} and computational mechanics~\cite{stefanos2015neural,kirchdoerfer2016data,bock2019review,han2019uniformly,huang2020learning,xu2020learning,masi2020thermodynamics} in general.

Most existing machine-learning-based constitutive models
are local, algebraic functions of the form $\tau = f(\mathbf{u})$, i.e., the closure variable (denoted as $\tau$) at $\bm{x}_i$ depends solely on the resolved variable (e.g., velocity $\mathbf{u}$) or at most the derivatives thereof (e.g., the strain rate) at the \emph{same} point $\bm{x}_i$. Such local neural network models are based on the commonly used turbulence models such as the eddy-viscosity models, which are derived based on the weak equilibrium assumption and assume a  linear or nonlinear mapping from the local mean strain-rate to the Reynolds stress anisotropy at the same point~\cite{strofer2021end}.
However, the underlying unresolved physics to be represented by the constitutive models can be nonlocal. This is most evidently seen in turbulence models. The unresolved turbulent velocity fluctuation (which can be described by its second order statistics, i.e., the correlation, Reynolds stress) depends on an upstream neighborhood due to the convection and diffusion~\cite{gatski1996simulation}. In fact, the transport physics of Reynolds stresses can be exactly described by a set of coupled convection--diffusion PDEs, which unfortunately contain unclosed terms of even higher order statistics. Such unclosed terms include triple velocity correlation, pressure-strain-rate and dissipation, all of which require extra modeling. The pressure--strain-rate term, in particular, greatly influences the performance of the Reynolds stress model but is notoriously difficult to model. Moreover, it is often challenging to achieve convergence in industrial applications due to unstructured grids of low quality and complex flow configurations. Therefore, despite the theoretical superiority, the Reynolds stress model has not been widely used because of its lack of robustness~\cite{basara2003new}. Other PDE-based constitutive models in industrial CFD include transport equations for intermittency~\cite{menter2015one} and amplification factor~\cite{coder2014computational} in laminar-turbulent transition modeling and the transport of eddy viscosity~\cite{spalart92one}, turbulent kinetic energy, dissipation~\cite{launder74application}, and frequencies~\cite{wilcox88reassessment} in turbulence modeling. Such closure models all imply a region-to-point mapping from the resolved primary variable (typically the flow field) to the respective closure variable.

In light of the observations above, it seems natural to explore using neural networks to mimic such nonlocal, region-to-point mapping for constitutive modeling. 
Such neural-network-based constitutive models both preserve the nonlocal mapping dictated by the underlying physics and admit diverse sets of training data.
A natural first attempt is to draw inspirations from image recognition, where convolutional neural networks (CNN) are used to learn a nonlocal mapping from a region of pixels to its classification (e.g., nature of the object it represents, presence of tumor). This avenue has been explored in our earlier work~\cite{zhou2020nonlocal}. 
We have demonstrated that the CNN-based model learns not only an accurate closure model but also the underlying Green's function when the physics is described by a linear convection--diffusion PDE. This observation was also confirmed in recent works from other groups~\cite{gin2020deepgreen,li2020neural}.

Unfortunately, constitutive modeling (and physical modeling in general) poses much more stringent requirements than those in computer vision applications, and a straightforward application of those methods would not suffice.
More specifically, the constitutive model must be \emph{objective}, i.e., indifferent to the material frame~\cite{speziale1998review}.  To put it plainly, a constitutive model should not vary depending on the coordinate system (e.g., origin and orientation) or reference systems for zero velocity (i.e., Galilean invariance) or pressure (absolute versus gauge pressure). The frame-independence requirement immediately rules out CNN as an ideal vehicle for general constitutive modeling except for the special problems that are already equipped with intrinsic coordinates (e.g., the normal and tangential directions of the wall in the modeling of near-wall turbulence). Needless to say, all equations in classical mechanics from Newton's second law to Navier--Stoke equations already satisfy the frame-independence requirement as they faithfully describe the physics, which are frame-independent. When developing constitutive models, the frame-independence requirement restricts the set of  functions that are admissible as constitutive models~\cite{spalart15philosophies}, be it stress--strain model for elastic materials, eddy viscosity models of turbulence~\cite{pope1975more,gatski1993on}, or pressure--strain-rate models for Reynolds stress transport equations~\cite{speziale1991modelling}. However, in the realm of data-driven modeling, such conventional wisdom is not directly applicable, as the candidate function forms are dictated by the adopted type of machine learning model. Specifically, as neural networks are successive compositions of linear functions and nonlinear squashing, it is not straightforward to restrict the function forms as practiced in traditional constitutive modeling~\cite{pope1975more,gatski1993on,speziale1991modelling}.
Recently, global operators or surrogate models parameterized by machine learning models for approximating solutions of the PDEs~\cite[see, e.g.,][]{long2018pde,long2019pde,sun2020surrogate,kim19deep,guo2016convolutional,lu2019deeponet,ma2020machine,ribeiro2020deepcfd,li2020neural,li2020multipole,li2020fourier} have emerged as a new computation tool, which also hold the promise to serve as nonlocal constitutive modeling. However, the objectivity of these modeling approaches, such as frame-independence and permutational invariance introduced below, has rarely been discussed in the literature.

Alternative strategies have been proposed to achieve frame-independence for data-driven modeling. The straightforward method borrowed from computer vision~\cite{krizhevsky2012imagenet,sharif2014cnn} is to augment the training data by duplicating them to different coordinate systems (e.g., rotated by a series of random angles) before conducting the training~\cite{ling16machine}. That way, functions that deviate from frame-independence too much are implicitly rejected in the training process. However, this method is inefficient and does not guarantee strict frame-independence.
Another strategy is to approximate the constitutive model in an admissible form, a function only processing invariants of the raw features, such as the magnitude of velocities, eigenvalues of strain-rate tensor~\cite{ling16machine}, Q criteria of the velocity~\cite{wang17physics-informed}, or other handcrafted scalars.
%\todo{I try to make the logic of this sentence clearer. To me, Q criteria is also an invariant of the velocity gradient tensor.}
However, while it does eliminate dependence on the coordinate orientation,  this may result in information loss on the relative direction among the input vectors (and tensors). For example, consider a hypothetical constitutive model where the turbulent kinetic energy $\tau$ (a scalar quantity) at a given location $\bm{x}_0$ is formulated as a function of the mean velocities $\mathbf{u}_1$, $\mathbf{u}_2$, and $\mathbf{u}_3$ at three cells in the neighborhood of $\bm{x}_0$.
Restricting admissible functions to the form $\tau = f(\vnorm{\mathbf{u}_1}, \vnorm{\mathbf{u}_2}, \vnorm{\mathbf{u}_3})$, i.e., depending only on the  velocity magnitudes, is too restrictive, although it does ensure rotational invariance. Rather, the \emph{pairwise} inner-products (mutual projections) of the velocities also need to be included as input to make input information complete~\cite{bartok2013representing,weyl1946classical} and still rotational invariant, i.e., $\tau = f(\vnorm{\mathbf{u}_1}, \vnorm{\mathbf{u}_2}, \vnorm{\mathbf{u}_3}, \mathbf{u}_1^\top \mathbf{u}_2, \mathbf{u}_1^\top \mathbf{u}_3, \mathbf{u}_2^\top \mathbf{u}_3)$. 
Note that our convention in this paper is that all vectors are column vectors.
More concisely it can be written as $\tau = f\big(\cR \cR^\top\big)$ with $\cR = \vstack{\mathbf{u}_1^\top, \mathbf{u}_2^\top, \mathbf{u}_3^\top}$ and $\vstack{\cdot}$ indicating vertical concatenation of velocity vectors to form matrix $\cR$.

In addition to the frame-independence that leads to translation and rotation invariance, the nonlocal mapping used in practice also involves another kind of invariance, the permutation invariance.
In computational mechanics, unstructured meshes are often employed for discretization (e.g., in finite volume or finite element methods), where there is no intrinsic order for indexing the cells.
Therefore, if we aim to construct an objective nonlocal mapping from the data in these cells, the mapping must be independent of the index order.
In the previous hypothetical example, this requirement dictates that the prediction $\tau$ must remain the same whether the input is  $(\mathbf{u}_1, \mathbf{u}_2, \mathbf{u}_3)$, $(\mathbf{u}_3, \mathbf{u}_1, \mathbf{u}_2)$, or any other ordering of three velocity vectors. In other words, the closure variable $\tau$ must only depend on the set as a whole and not on the ordering of its elements.

In various fields of physical modeling, it has been observed that neural-network (and other data-driven models) that respect all the physical constraints and symmetries tend to achieve the best performance~\cite{e2020integrating,wu2019representation,zafar2020convolutional,taghizadeh2020turbulence,doan2020autoencoded}. In light of such insight, in recent years researchers have made significant efforts in to imposing physical constraints on neural networks for emulation,  closure modeling, reduced-order modeling and discovering of physical systems~\cite[see, e.g.,][]{wu2018physics-informed,wu2020enforcing,yang2019predictive,doan2020physics,doan2021short,doan2020autoencoded,schmelzer2020discovery,yu2020data}.
In the present work we draw inspirations from the recently developed neural-network-based potential, Deep Potential~\cite{han2018deep,zhang2018end}, to develop the nonlocal closure model that satisfies the aforementioned invariant properties in the context of computational mechanics.
Potential energy is the basic but central building block in molecular dynamics, which are common tools in many disciplines, including physics, chemistry, and material science. 
Potential energy is a scalar function that takes all the considered atoms' positions and types as input. Physics dictates that the potential energy is frame-independent and permutational invariant with atoms of the same type.
Deep Potential guarantees all these invariances and has achieved remarkable success in accelerating the simulation of large systems with \textit{ab initio} accuracy~\cite{zhang2018deep,zhang2018deepcg,jia2020pushing} thanks to the approximating ability of neural networks.

\begin{figure}
    \centering
    \includegraphics[width=1\textwidth]{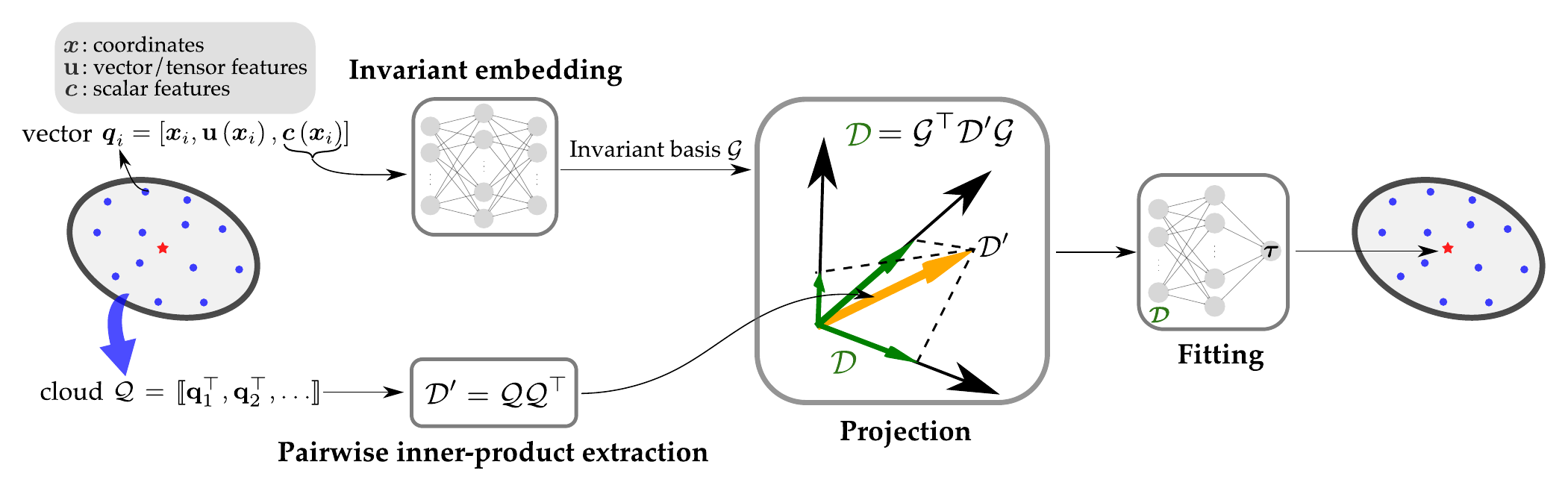}
    \caption{
    Schematic of the frame-independent, permutation-invariant vector-cloud neural network for nonlocal constitutive modeling, showing a mapping $\cQ \mapsto \tau$ from a cloud (left oval) of feature vectors $\cQ = \vstack{\bq_1^\top, \bq_2^\top, \ldots}$ to the closure variable $\tau$ (center of right oval).
    We construct the mapping by starting with two simultaneous operations: (i) extract pairwise inner-product to obtain rotational invariant features $\cD'_{ii'} = \bq_i^\top \bq_{i'}$ and (ii) map the scalar quantities $\bw$ in each vector $\bq$ in the cloud through an embedding network to form a permutational invariant basis $\cG$, which also inherits its rotational invariance from input~$\bw$. Then, we project $\cD'$ onto basis $\cG$ (not necessarily orthogonal) to produce final feature matrix $\cD$, which is invariant to frame and permutation. Finally, we fit a neural network to map features $\cD$ to closure variable $\tau$. More details on the workflow and implementation are shown in Fig.~\ref{fig:NN-architecture}.
    \label{fig:framework}
    }
\end{figure}

Our main contribution in the present work is in using a frame-independent \emph{vector-cloud neural network} to capture the nonlocal physics in convection--diffusion PDEs, which are commonly used in closure modeling. The vector-cloud network maps the input data matrix $\cQ$, defined by a set of points in a region (referred to as \emph{cloud}), to the closure variable $\tau$ as shown in Fig.~\ref{fig:framework}. Each point in the cloud has a vector $\bq$ attached to it, which defines a row in $\cQ$. The vector consists of the point's frame-dependent coordinates $\bx$ and resolved variables $\bu$ (e.g., velocity or strain) as well as frame-independent scalar features $\bw$.
The input features of the vector-cloud network (detailed in Table~\ref{tab:features} later) are specifically tailored to reflect the underlying transport physics, but the network is otherwise similar to Deep Potential.
The vector-cloud neural network is constructed to be invariant both to coordinate translation and rotation and to the ordering of the points. The invariance is achieved by the following operations as illustrated in Fig.~\ref{fig:framework}:
(i) compute pairwise inner products $\cD'_{ii'} = \bq_i^\top \bq_{i'}$ among all vectors in the cloud to obtain rotational invariant features, (ii) map the scalar features to a rotational invariant basis $\cG$ with an embedding network, and (iii) project the inner product $\cD'$ onto the basis to form invariant feature $\cD = \cG^\top \cD' \cG$, which is then further mapped to $\tau$ with a neural network.
The mapping $\cQ \mapsto \tau$ is frame-independent and permutational invariant as $\cD$ is invariant. More details of the workflow will be presented in Section~\ref{sec:method} and further illustrated in Fig.~\ref{fig:NN-architecture}.

\section{Problem statement and proposed methodology} 
\label{sec:method}
In the context of turbulence modeling for incompressible flows of constant density $\rho$, the mean flow fields (velocity $\mathbf{u}$ and pressure $p$) are described by the following RANS equations:
\begin{equation}
    {\mathbf{u}} \cdot \nabla {\mathbf{u}} - \nu \nabla^2 \mathbf{u} = -\frac{1}{\rho} \nabla {p} + \nabla \cdot \bm{\tau},
\end{equation}
where $\nu$ is molecular viscosity and
the Reynolds stress term $\bm{\tau}$ needs to be modelled by a transport equation of the form:
\begin{equation}
    \label{eq:rstm}
    \mathbf{u} \cdot \nabla \bm{\tau} - \nabla \cdot (\nu \nabla \bm{\tau})  = \mathsf{P} - \mathsf{E},
\end{equation}
where $\mathsf{E}$ indicates dissipation and $\mathsf{P}$ includes both production, turbulent transport, and pressure--strain-rate terms. The constitutive PDE~\eqref{eq:rstm} implies a nonlocal mapping $g: \mathbf{u}(\bm{x}) \mapsto \bm{\tau}(\bm{x})$ from the mean velocity field to the Reynolds stress. Our overarching goal of the data-driven constitutive modeling is to build a neural-network-based surrogate for the equation~\eqref{eq:rstm} above that can be trained and used more flexibly with diverse sets of data. 

In this preliminary work we prove the concept on a problem that is closely related yet simplified in two aspects. First, we build a constitutive neural network for a scalar $\tau$ as opposed to the Reynolds stress tensor $\bm{\tau}$. Examples of scalar quantities in turbulence modeling include turbulent kinetic energy (TKE) and turbulent length scale. In this work, we consider a hypothetical, dimensionless concentration tracer (e.g., volume fraction in multiphase flows) as a scalar $\tau$.
Second, the scalar $\tau$ is transported by a velocity field~$\bu$ not affected by the scalar and obtained from laminar flow. 
Given these two simplifications, hereafter $\tau$ can be interpreted as the concentration field of a passive tracer, determined by the resolved variable (velocity~$\mathbf{u}$).
The transport equation otherwise mimics Eq.~\eqref{eq:rstm} above and reads:
 \begin{align}
    \label{eq:scalar}
    & \bu \cdot \nabla \tau - \nabla \cdot (C_\nu \nabla \tau)  = \mathsf{P} - \mathsf{E} \\
    \text{with} \quad
    & \mathsf{P}  = C_g  \ell_{m}  \sqrt{\tau}  s^2  \; \text{and} \;
    \mathsf{E} =  C_\zeta \tau^2 \notag \\
    & s = \| \mathbf{s}\| = \tnorm{\nabla \mathbf{u} + (\nabla \mathbf{u})^\top} ,
    \notag
\end{align}
where the production $\mathsf{P} $ depends on scalar $\tau$, mixing length $\ell_{m}$, and strain-rate magnitude $\tnorm{\mathbf{s}}$. This transport equation resembles that for the TKE transport equation~\cite{pope00turbulent} (see \ref{app:tke} for detailed analogy between the two equations).
The mixing length is proportional to wall distance $\hat{\eta}$ and capped by the boundary layer thickness $\delta$, i.e., 
\begin{equation}
\ell_{m} = \min(\kappa \hat{\eta}, C_\mu \delta)
\label{eq:ellm}
\end{equation}
with von Karman constant $\kappa = 0.41$ and coefficient $C_\mu = 0.09$. Finally, $C_g=200$ \si{m^{-1} s}, $C_\nu=0.1$ \si{m^2/s}, and $C_\zeta = 3$ \si{s^{-1}} are coefficients associated with production, diffusion, and dissipation, respectively. 
Although the PDE above is in a dimensional form with physical units, the inputs of the vector cloud neural network will exclusively use dimensionless quantities (see Table~\ref{tab:features} later). As such, 
training and prediction can be performed in flows of vastly different scales, as long as they are dynamically similar. Such a dynamic similarity is achieved if the coefficients $C_\mu$, $C_g$, and $C_\zeta$ are properly scaled according to the length- and velocity-scales of the system. Details of the similarity are discussed in \ref{app:nondimensionalize}.

\begin{figure}[!htb]
\centering
\includegraphics[width=1.0\textwidth]{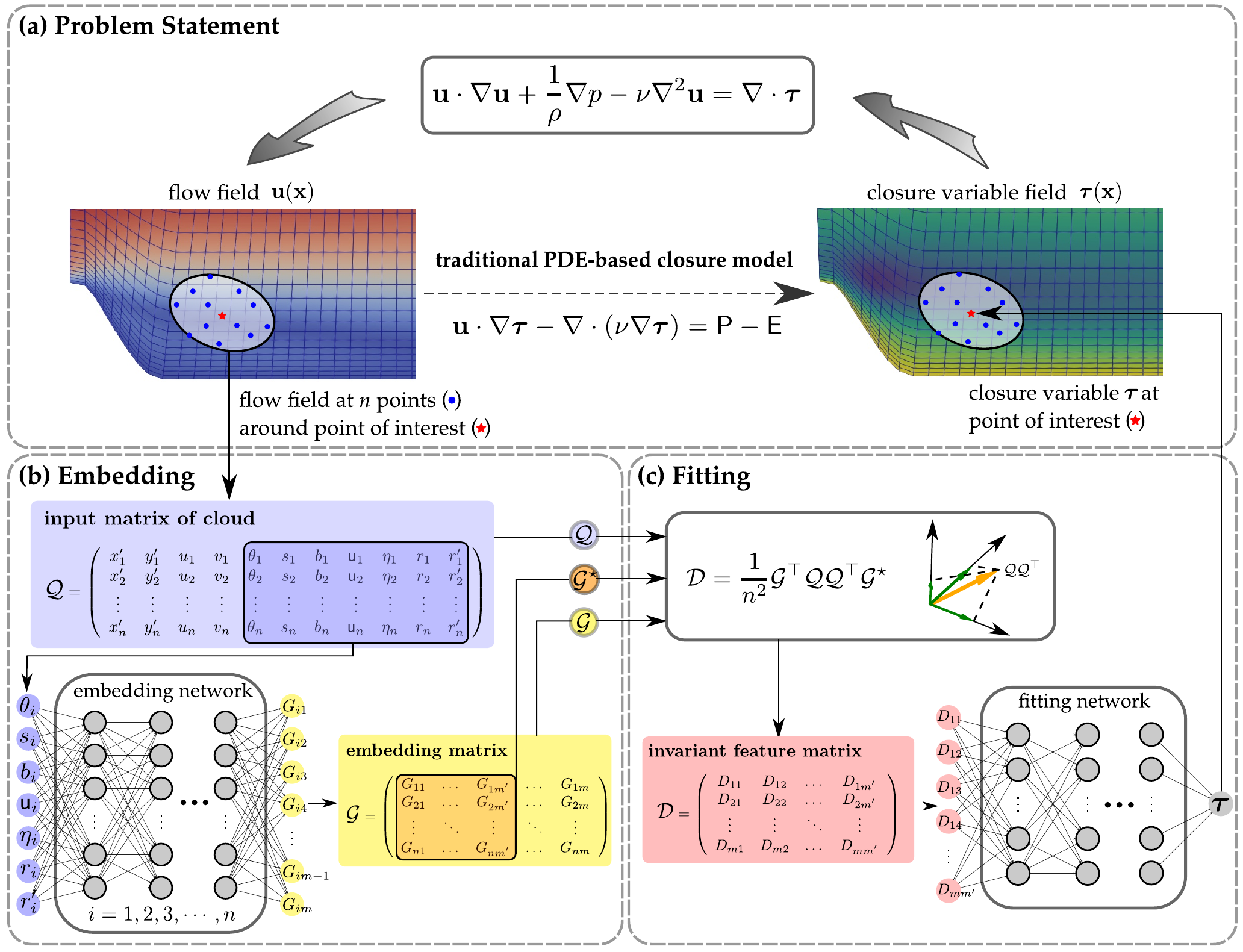}
  \caption{
  Detailed schematic of the vector-cloud neural network to provide nonlocal closure model $\bm{\tau}(\bu$) for the primary equation (the Reynolds-averaged Navier--Stokes equation is used here for illustration): (a) generate labeled training data ${(\mathcal{Q}, \bm{\tau})}$, where the input matrix $\mathcal{Q} \in \mathbb{R}^{n \times l}$ consists of $n$ vectors, each attached to a point ($\color{blue} \bullet$) in the cloud (\protect\tikz{ \protect\draw (0,-5) ellipse (6pt and 2.5pt);}) surrounding the point ($\color{red} \star$) where the closure variable $\bm{\tau}$ is to be predicted; each vector (a row in $\cQ$) encodes its relative coordinate $\bx'$, velocity $\bu$, and other scalar features (Table~\ref{tab:features}); (b) map the scalar features in $\mathcal{Q}$ through an embedding network to a set of invariant bases $\mathcal{G} \in \mathbb{R}^{n \times m}$, of which $\mathcal{G}^\star \in \mathbb{R}^{n \times m'}$ is the first $m'$ ($\le m$) columns; then, project the pairwise inner-product matrix $\mathcal{Q} \mathcal{Q}^\top$ to the learned embedding matrix $\mathcal{G}^\top$ and  its submatrix $\mathcal{G}^\star$ to yield an invariant feature matrix $\mathcal{D} \in \mathbb{R}^ {m \times m'}$; and (c) flatten and feed the feature matrix $\mathcal{D}$  into the fitting network to predict the closure variable $\bm{\tau}$. The constitutive mapping $\mathbf{u(\bx)} \mapsto \bm{\tau}$ based on the vector-cloud neural network is invariant to both frame translation and rotation and to the ordering of points in the cloud.
  \label{fig:NN-architecture}
  }
\end{figure}

Given the problem statement, we construct a constitutive neural network to predict the tracer concentration $\tau$ at the point of interest $\bm{x}_0$ from the flow field information.
Considering both the nonlocal physics embodied in the transport PDE~\eqref{eq:scalar} and feasibility for implementation in CFD solvers, the network should form a region-to-point mapping $\hat{g}: \region{\mathbf{q}(\bm{x}_0)} \mapsto \bm{\tau}(\bm{x}_0)$, where $\mathbf{q}$ is the feature vector, and $\region{\mathbf{q}(\bm{x}_0)}$ indicates the collection of features $\{\bq_i\}_{i=1}^n$ on $n$ points sampled from the region around $\bm{x}_0$ (e.g., cell centers in the mesh indicated in dots inside the ovals in Fig.~\ref{fig:NN-architecture}a).
In general, the number of points $n$ in a cloud can vary from location to location.
The feature vector $\mathbf{q}$ at each point is chosen to include the relative coordinate $\bm{x}' = \bm{x} - \bm{x}_0$, flow velocity $\bu$, and additional seven scalar quantities $\bw = [\theta, s, b, \eta, \mathsf{u}, r, r']^\top$,
including
\begin{enumerate}[(1)]
    \item cell volume $\theta$, which represents the weight for each point in grid-based methods;
    \item velocity magnitude~$\mathsf{u}$ and strain rate magnitude~$s$, the latter of which often appears in various turbulence and transition models;
    \item boundary cell indicator $b$ and wall distance function~$\eta$, which help the closure model to distinguish between PDE-described mapping and wall model (boundary condition);
    \item proximity~$r$ (inverse of relative distance) to the center point of the cloud and proximity $\displaystyle r'$ defined based on shifting and normalization of projection $-\bu^\top \bm{x}'$ accounting for the alignment between the convection and the relative position of the point in its cloud.
    
    The former is motivated by the fact that features on points closer to $\bm{x}_0$ have more influences on the prediction $\tau(\bm{x}_0)$ than those further away for isotropic, diffusion-dominated physics. The latter is justified as the velocity-relative position alignment is important for convective physics: a cell with velocity pointing towards the cloud center $\bm{x}_0$ is more important than those with velocities perpendicular to or pointing away from the center.
    Hence we propose $r'$ to be proportional to (i) the time scale $\frac{\vnorm{\bu}}{\vnorm{\bx'}}$ and (ii)
    the cosine of the angle (shifted approximately to the positive range [0, 2]) between the velocity $\bu$ and the vector $-\bm{x}' (=\bm{x}_0 - \bm{x})$ from the point to the cloud center, i.e., $\epsilon_1-\frac{\bu^\top \bm{x}'}{\vnorm{\bu} \vnorm{\bx'}+\epsilon_2}$ with $\epsilon_1 = 1.05$ and $\epsilon_2 = 10^{-10}$ \si{m^2/s}. 
    
\end{enumerate}
Note that $s$ and $\mathsf{u}$ are directly derived from the velocity $\mathbf{u}$, and the features $r$ and $r'$ can be derived from the relative coordinates (and velocities) of the points on the cloud. While including these derived quantities introduces some redundancy, it helps to guide the construction of the network-based closure model by embedding prior knowledge of physics into the input. 
We also include two scalar quantities in the input features to represent the boundary conditions: (1) boundary cell indicator $b$ and (2) wall distance function $\eta$, both of which reflect the position relative to the wall boundaries.
The feature engineering above leads to $\mathbf{q} =\vstack{\bm{x}', \mathbf{u}, \bm{c}}$. The relative coordinates $\bx'=[x',y']^\top$ and velocity $\bu=[u,v]^\top$ are in two dimensional spaces ($d = 2$), but extension to three dimensions is straightforward. With $l' = 7$ scalar quantities as chosen above, the feature vector is $\bq \in \bbR^l$ with $l = 2d+l' = 11$.
The features and the associated pre-processing are presented in Table~\ref{tab:features}.
Note that we introduce the characteristic length $L_0$, flow velocity $U_0$ and time $T_0={L_0}/{U_0}$ to non-dimensionalize the features such that all the features are dimensionless before being fed into the neural network, which enables the output of the neural network to be comparable even in systems with different scales. In this work, we take the height of the hill as the characteristic length $L_0=H=1$ m and the mean bulk velocity at the inlet patch as the characteristic velocity $U_0=\left|\mathbf{u}_{b}\right|=1$ m/s. The characteristic time is then calculated as $T_0 = {L_0}/{U_0} = 1$ s. 
By the definition in Table~\ref{tab:features}, the relative coordinate features $\bx'=[x',y']^\top$ are essentially the cosine and sine of the angle between the relative direction and the streamwise direction, and thus should be more precisely referred to as \emph{relative direction}. However, hereafter it is referred to as (normalized) ``relative coordinates" according to the raw features. 

\begin{table}[!htb]
    \caption{Flow features vector $\mathbf{q}$ used as input to the neural network arranged in three groups separated by lines: relative coordinates, velocities, and scalar quantities, i.e., $\mathbf{q} = \vstack{\bx', \bu, \bw}$. The final input features are obtained by processing the corresponding raw features. Notations are as follows:
    $\bm{x}_0 = (x_0, y_0)$, coordinate of cloud center;
    $\bx' = \bx - \bx_0$, relative coordinate to cloud center, 
    $\epsilon_0 (=10^{-5})$, a small number to avoid singularity; $\hat{u}$ and $\hat{v}$, $x$- and $y$-component of velocity $\bu$; $\overline{\theta}$, the average cell volume for all sampled data points within a cloud; $\epsilon_r = 0.01$,  constant in reciprocal to ensure the diminishing importance of point towards the edge of the cloud;
    $\eta$, the wall distance normalized by boundary layer thickness scale $\ell_\delta$ and capped by 1; 
    $\epsilon_{1}=1.05, \epsilon_2=10^{-10}$ \si{m^2/s} are used to shift the cosine of the alignment angle to positive range;
    {$\|\,{\cdot}\,\|$} and $\vnorm{\cdot}$ indicates tensor and vector norms. $L_0$, $U_0$ and $T_0$ denote the characteristic length, flow velocity and time, respectively.
    }

    \centering
    \begin{tabular}{P{0.13\textwidth} | P{0.1\textwidth}  P{0.37\textwidth} P{0.2\textwidth}}
    \toprule
    \textbf{category} & \textbf{feature} &  \textbf{description of raw feature}  & \textbf{definition} \\ 
    \midrule
    \multirow{2.2}{*}{\begin{tabular}{c}relative \\coordinates\end{tabular}} &  $x'$  & relative $x$-coordinate to cloud center & \(\displaystyle\frac{{x - x_{0}}}{\vnorm{\bm{x} - \bm{x}_{0}}+\epsilon_0}\)  \\ 
    & $y'$  &  relative $y$-coordinate to cloud center & \(\displaystyle\frac{{y - y_{0}}}{\vnorm{\bm{x} - \bm{x}_{0}} + \epsilon_0}\)  \\ 
    \midrule 
    \multirow{2}{*}{\begin{tabular}{c}velocity \\vector\end{tabular}} & $u$  &  $x$-component of velocity & $\hat{u}/U_0$ \\
    &  $v$  & $y$-component of velocity  &  $\hat{v}/U_0$  \\ 
    \midrule 
    \multirow{7.5}{*}{\begin{tabular}{c}scalar \\quantities\end{tabular}} &  $\theta$ & cell volume $\hat{\theta}$ &  $\hat{\theta} / \overline{\theta}$ \\
    & $s$  &  magnitude of strain rate $\mathbf{s}$  & $\|\mathbf{s}\| T_0$   \\ 
    & $b$  &  boundary cell indicator   & 
    1 (yes) or 0 (no) 
    \\
    &  $\mathsf{u}$  & velocity magnitude  & $\vnorm{\bu}/U_0$  \\ 
    &  $\eta$  & wall distance $\hat{\eta}$  &   
    $\min(\hat{\eta}/\ell_\delta, 1)$  \\
    &  $r$  & proximity to cloud center & $\displaystyle{\frac{\epsilon_r}{\vnorm{\bm{x} - \bm{x}_{0}}/L_0 + \epsilon_r}}$ \\
    &  $r'$ & proximity in local velocity frame &  $\displaystyle{\frac{r\vnorm{\bu}}{U_0}\big( \epsilon_1-\frac{\bu^\top \bm{x}'}{\vnorm{\bu} \vnorm{\bx'}+\epsilon_2}\big)}$ \\
    \bottomrule
    \end{tabular}
    \label{tab:features}
\end{table}

In light of the frame-independence requirements, our first attempt is to seek a mapping of the form 
$\tau = \hat{g}(\cR \cR^\top)$ with $\cR = [\mathbf{q}_1, \ldots, \mathbf{q}_n]^\top$.
Such a function has both translational and rotational invariance, because it depends only on the relative coordinates $\bm{x}_i'$  and relative orientations of $\bq_i$ in the form of pairwise projections $\cR \cR^\top$ among the feature vectors. However, it lacks permutational invariance as it depends on the numbering (ordering) of the $n$ cloud points. This is more evident when we write the projection in index form $\cD'_{i i'} = \sum_{j=1}^{11} \cR_{ij} \cR_{ji'}$, where $i$ and $i'$ are cloud point indices, and $j$ is feature index.  Matrix $\cD'$ would switch rows and columns if the $n$ points are permutated (i.e., the numbering is switched). Therefore, we need to define a function that  depends only on the set of vectors $\{\bq_i\}_{i=1}^n$ and not on its ordering. To this end, we introduce a set of $m$ functions $\{\phi_k(\bw_i)\}_{k=1}^m$ for the scalar quantities $\bw_i$ (note the input $\bw_i = [\theta_{i}, s_i, \cdots]^\top$ represents all the scalar quantities for the $i^\text{th}$ point, which is part of the feature vector $\bq_i$ and already has translational and rotational invariance) and define $\cL_{k j} = \frac1n\sum_{i=1}^n \phi_k(\bw_i) \, \cR_{ij}$, where $k=1, \ldots, m$ is the function index. The summation over the point index~$i$ removes the dependence of $\cL$ on the ordering and make it permutational invariant. If we define a matrix $\cG_{i k} = \phi_k(\bw_i)$, the order-removing transformation above can be written in matrix form as $\cL=\frac{1}{n} \cG^\top \cR$, where the normalization by $n$ allows training and prediction to use different numbers of sampled points in the cloud. This mapping $f_\text{embed}: \cC \mapsto \cG$ with $\cC = \llbracket \bw_1^\top, \bw_2^\top, \ldots, \bw_n^\top \rrbracket$ is implemented as an embedding neural network with $m$-dimensional output, interpreted as applying the $m$ functions $\{\phi_k(\bw_i)\}_{k=1}^m$ to the frame-independence part $\bw_i$ of the feature vector $\{\bq_i\}$ at point~$i$. This is illustrated in Fig.~\ref{fig:NN-architecture}b. Similarly, we can define $\cL^\star = \frac{1}{n} \cG^{\star\top} \cR$, with $\cG^{\star}$ being the first $m'$ columns of $\cG$, and $\cL^\star$ is also permutational invariant. 
Here we choose $\cG^{\star}$ as a subset of $\cG$ rather than $\cG$ itself mainly for the purpose of saving the computational cost without sacrificing the accuracy.
Next, we define a projection mapping $\cD = \cL \cL^{\star\top} = \frac{1}{n^2}\cG^\top\cR\cR^\top\cG^\star$, then $\cD$ has translational, rotational, and permutational invariance (since both $\cL$ and $\cL^{\star\top}$ have permutational invariance while the pairwise projection $\cL \cL^{\star\top}$ achieves rotational invariance);
it retains all information from the feature vectors on the cloud via $\cR\cR^\top$ by projecting it onto the learned basis $\cG$.
Finally, we fit a function $f_\text{fit}: \cD \mapsto \tau$ that maps $\cD$ to the closure variable $\tau$, which is achieved through a fitting network shown in Fig.~\ref{fig:NN-architecture}c. The combination of the embedding network, the linear mapping, and the fitting network thus forms the complete constitutive network that serves as the closure for the primary equation. 

In summary, the data-driven, frame-independent, nonlocal constitutive model $\hat{g}: \cR \mapsto \tau$ that maps the features for points on a cloud to the closure variable is achieved in the following four steps: 
\begin{enumerate}[(1)]
    \item Feature engineering to stack relative coordinate $\bx'$, velocity $\bu$, and scalar quantities $\bw$ (each row for a point in cloud):
    \begin{equation}
    \cR = \begin{bmatrix}
         \mathbf{q}_1^\top \\
          \mathbf{q}_2^\top  \\
          \vdots  \\
           \mathbf{q}_n^\top 
    \end{bmatrix} = \left[
 \begin{array}{cc|cc|ccccccc}
x'_{1} & y'_{1}  & u_{1}  & v_{1} & \theta_1 & s_{1} & b_{1} & \mathsf{u}_{1} & \eta_{1} & r_{1} & r'_{1}   \\
x'_{2} & y'_{2} & u_{2}  & v_{2} & \theta_2 & s_{2} & b_{2} & \mathsf{u}_{2} & \eta_{2} & r_{2}& r'_{2}  \\
 \vdots & \vdots & \vdots & \vdots & \vdots & \vdots & \vdots & \vdots & \vdots & \vdots & \vdots\\
 x'_{n} & y'_{n} & u_{n}  & v_{n} &\theta_n & s_{n} & b_{n} & \mathsf{u}_{n} & \eta_{n} & r_{n} & r'_{n}
 \end{array}
\right] = [\cX, \cU, \mathcal{C}] \in \bbR^{n \times 11} ,
\end{equation}
where $\mathbf{q}^\top =[\bm{x}'^\top, \mathbf{u}^\top, \bm{c}^\top]$, and $\cX, \cU, \mathcal{C}$ correspond to the three submatrices of $\cR$ indicated by separators.
\item Embedding via a neural network $f_\text{embed}: \cC \mapsto \cG$ (using only scalar quantities of the feature vector) to remove the dependence on point numbering:
\begin{equation}
      \cG =\left[ \begin{array}{cccc|ccc}
         \phi_1(\bw_1) & \phi_2(\bw_1) & \ldots & \phi_{m'}(\bw_1) & \phi_{m'+1}(\bw_1) & \ldots   & \phi_m(\bw_1)\\
          \phi_1(\bw_2) & \phi_2(\bw_2) &  \ldots & \phi_{m'}(\bw_2) & \phi_{m'+1}(\bw_2) & \ldots  & \phi_m(\bw_2) \\
          \vdots        & \vdots         & \ddots & \vdots & \vdots & \ddots & \vdots  \\
           \phi_1(\bw_n) & \phi_2(\bw_n) & \ldots & \phi_{m'}(\bw_n) & \phi_{m'+1}(\bw_n) & \ldots  & \phi_m(\bw_n) 
    \end{array} \right] \in \bbR^{n \times m} ,
\end{equation}
and  $\cG^\star \in \bbR^{n\times m'}$ is the first  $m'$ ($\le m$) columns of $\cG$ (left of the separator line of $\cG$).
\item Combining $\cR \cR^\top$ with $\cG$ and $\cG^\star$ in a projection mapping to obtain a feature matrix with translational, rotational, and permutational invariance:
\begin{equation}
     \cD = \frac{1}{n^2}\cG^\top\cR\cR^\top\cG^\star
    \in \bbR^{m \times m'}.
\end{equation}
\item Fitting a neural network $f_\text{fit}$ to map the frame-independent input $\cD$ to closure variable $\tau$:
\begin{equation}
    \tau(\bx_0)  = f_\text{fit}\big(\cD).
\end{equation}
\end{enumerate}
In this work we choose $m=64$ embedding functions and a subset of $m'=4$ in the extraction of feature matrix $\cD$. The choice of $m,m'$ is similar to that in Deep Potential~\cite{zhang2018end}, which gives good performance in the prediction of the potential energy surface in practice. In our problem, we find the prediction performance is insensitive to $m'$ when $m'$ is much less than $m$. We set $m'=4$ because more features are preserved in the invariant feature matrix but the computational cost for the fitting network is still relatively small.
The problem statement of constructing the data-driven constitutive model and the complete procedure is summarized in Fig.~\ref{fig:NN-architecture}. A more general presentation of the framework and the proof of its invariance properties are presented in~\ref{app:proof}. Furthermore, a minimal example illustrating all the invariance properties is provided in~\ref{app:mini-example}.

\section{Results}
\label{sec:res}

\begin{figure}[!htb]
\centering
\subfloat[baseline geometry]
{\includegraphics[height=0.165\textwidth]{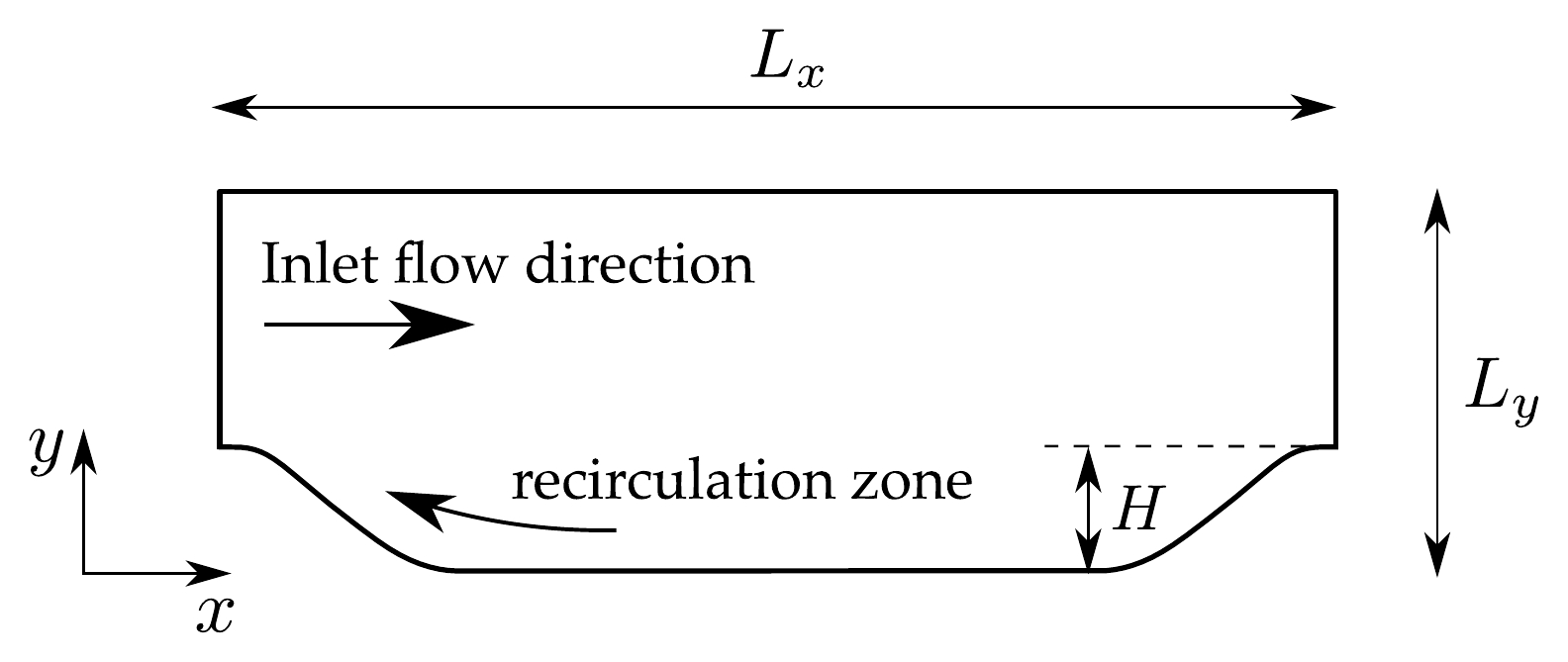}}
% \hspace{0.1em}
\subfloat[geometries of varying slopes]
{\includegraphics[height=0.165\textwidth]{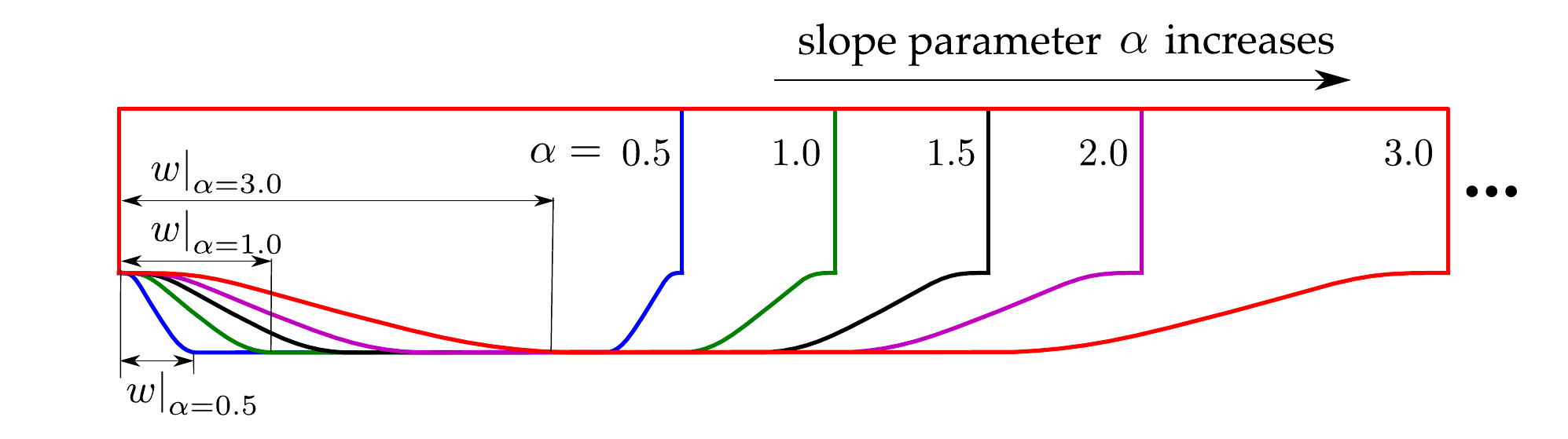}}
  \caption{
  Geometries of flow domains used for training and testing, showing (a) baseline geometry and (b) geometries with varying slopes, parameterized by the ratio $\alpha$ of the width $w$ of hill to that of the baseline hill, while the hill height $H=1$ m is the same for all geometries. 
  The $x$- and $y$-coordinates are aligned in the streamwise and wall-normal directions, respectively.
  By definition the baseline geometry has $\alpha=1$.
  The slope parameters $\alpha$ for training and testing are within the ranges $[1,\,2]$ and $[0.5,\,4]$, respectively.}
  \label{fig:geometry}
\end{figure}

In this study, we consider the flow over periodic hills. This case is representative as a benchmark for RANS simulations of flow separation due to the wall curvature~\cite{breuer2009flow}. 
The baseline geometry is shown in Fig.~\ref{fig:geometry}a with the hill shape described as piecewise polynomials. In the baseline geometry, the width and height of the hill are $w = 1.93$~m and $H = 1$~m, while the length and the height of the domain are $L_x = 9$~m and $L_y = 3.035$~m. The slope parameter $\alpha$ is varied to generate a family of geometries for training and testing~\cite{xiao2020flows}, which is presented in Fig.~\ref{fig:geometry}b. The parameter $\alpha$ describes the steepness of the hill profile, which equals the ratio of the width $w$ of the parameterized hill to that  of the baseline hill ($\left.w\right|_{\alpha=1}$). The hill height is kept the same for all geometries and thus the hill becomes more gentle with increasing $\alpha$. For training flows, we choose 21 configurations with $\alpha = 1.0, 1.05, \ldots, 2$. For testing flows, the parameters $\alpha$ range from 0.5 to 4. 

\begin{figure}[!htb]
\centering
{\includegraphics[width=0.48\textwidth]{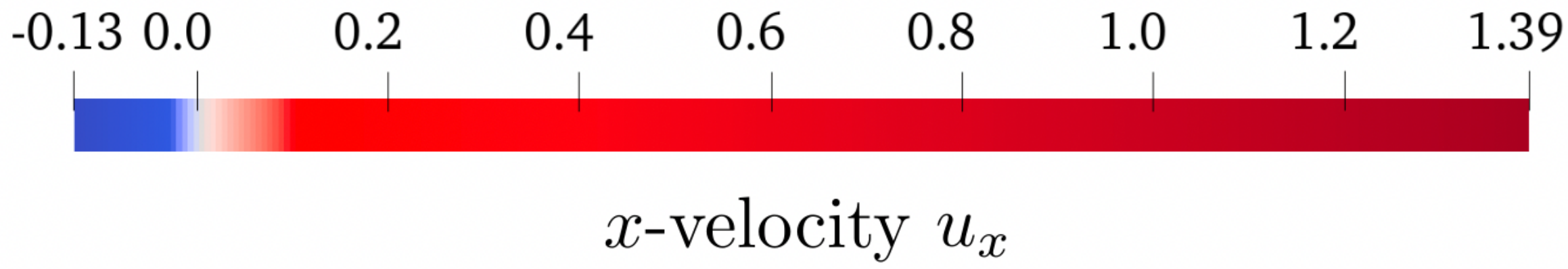}
\vspace{-6pt}}\\
\subfloat[$\alpha = 1$~(training, steepest)]
{\includegraphics[height=0.10\textwidth]{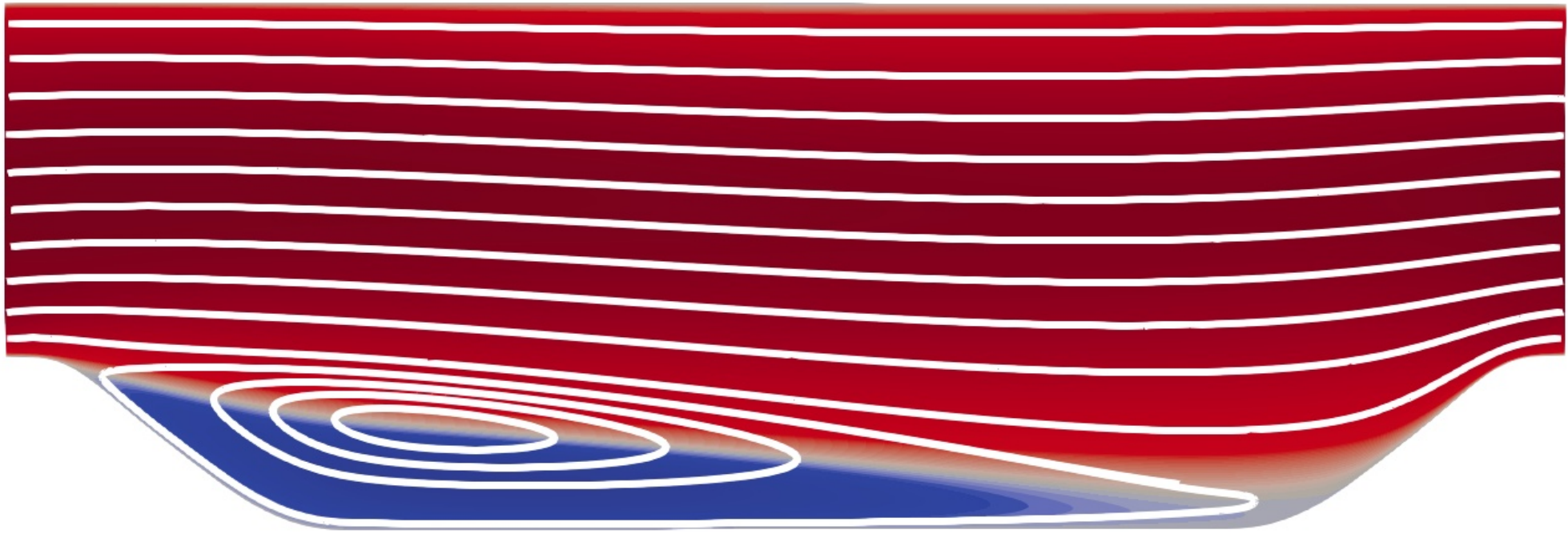}}
\hspace{0.1em}
\subfloat[$\alpha = 2$ (training, most gentle slope)]
{\includegraphics[height=0.10\textwidth]{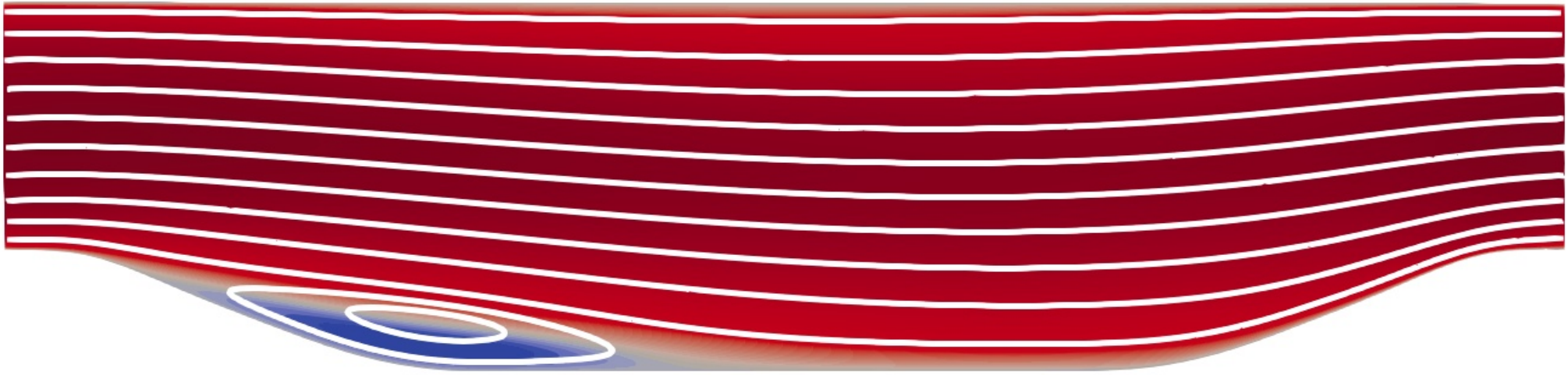}}\\
\subfloat[$\alpha=0.5$ (testing, steepest)]
{\includegraphics[height=0.10\textwidth]{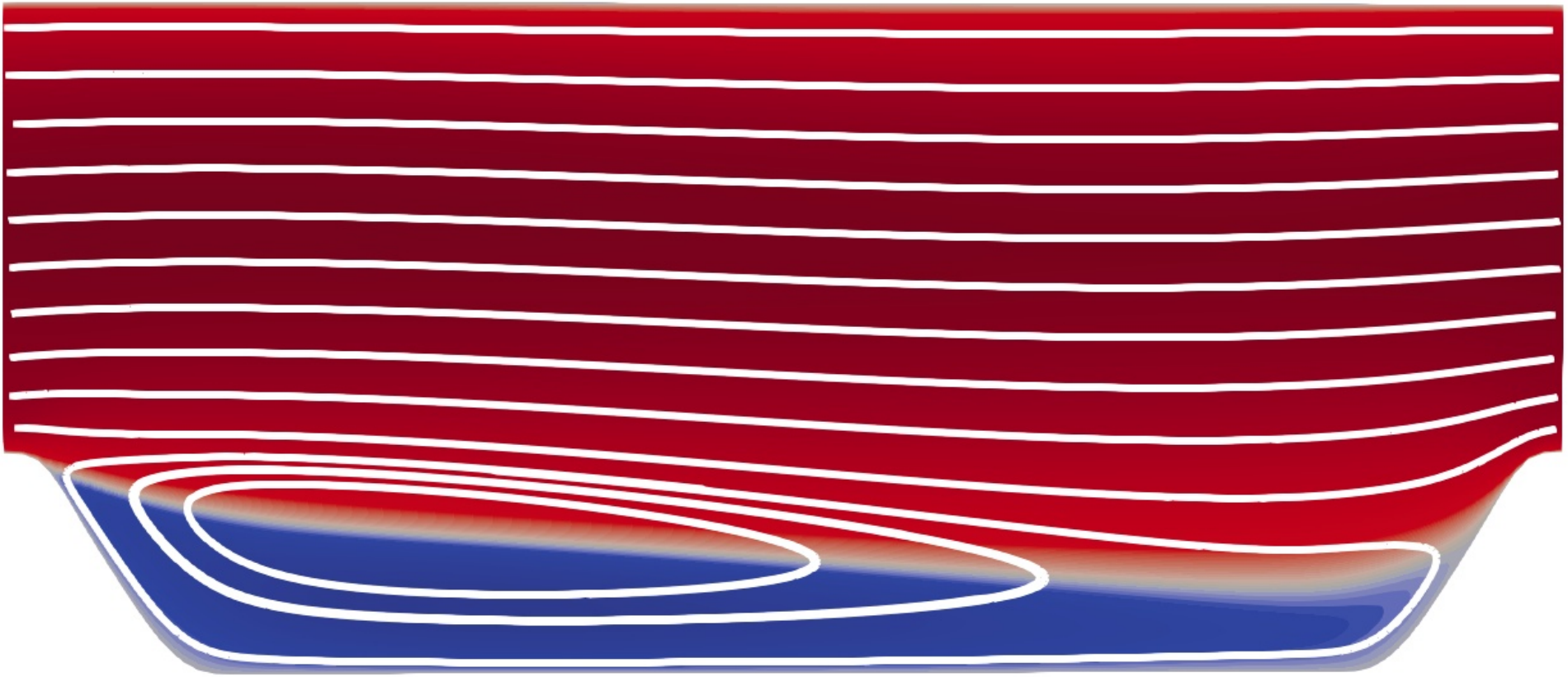}}
\hspace{0.1em}
\subfloat[$\alpha = 4$ (testing, most gentle slope)]
{\includegraphics[height=0.10\textwidth]{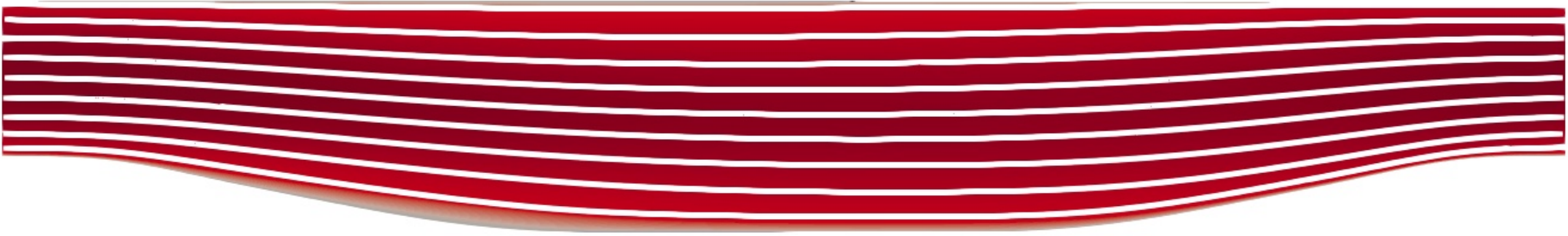}}\\
  \caption{Flow velocity fields used to solve the transport equation for generating data, showing streamlines from representative flow fields in the training and testing datasets. 
  Top: (a) the steepest slope $\alpha=1$ and (b) the most gentle slope ($\alpha=2$) in \textbf{training} flows. Bottom: (c) the steepest slope $\alpha=0.5$ and (d) the most gentle slope ($\alpha=4$) in \textbf{testing} flows, which are steeper and more gentle, respectively, than (a) and (b), the slope range of training flows. The color contours of streamwise velocity $u_x$ highlight the recirculation zones with blue/dark grey, showing the regions with flow reversal (present only in panels a--c).
  }
  \label{fig:flowfields}
\end{figure}

\subsection{Generation of training data}
We calculate the concentration field $\tau(\bx)$ by solving the transport PDE \eqref{eq:scalar} with a steady, laminar flow field $\mathbf{u}(\bx)$. The flow is driven by a constant pressure gradient such that the Reynolds number based on bulk velocity at the crest $|\bu_b|=1$~m/s and the hill height $H=1$~m reaches a specified value $Re = 100$.   All the numerical simulations are performed in the open-source CFD platform OpenFOAM~\cite{opencfd21openfoam}. We first simulate the steady-state flows over the periodic hill by solving the incompressible Navier--Stokes equations with the built-in solver simpleFoam. The coupled momentum and pressure equations are solved by using the SIMPLE (Semi-Implicit Method for Pressure Linked Equations) algorithm~\cite{issa86solution}.
Then, Eq.~\eqref{eq:scalar} is solved to simulate the steady-state concentration field $\tau(\bm{x})$ with the obtained stationary velocity field to provide data for training and testing as detailed below. When solving the fluid flow and concentration equations, at both walls nonslip boundary conditions ($\mathbf{u} = 0$) are used for the velocities and Dirichlet boundary conditions ($\tau = 0$) for the concentration, while cyclic boundary conditions are applied at the inlet and the outlet. The equations are numerically discretized on an unstructured mesh of quadrilateral cells with second-order spatial discretization schemes. In the wall normal direction all configuration has 200 cells that are refined towards both the upper and bottom walls.  In the streamwise direction the number of cells is approximately proportional to the length $L_x$ of the domain (200 cells in the baseline geometry with $\alpha=1$ and $L_x = 9$~m). The boundary layer thickness length scale is set to $\ell_\delta = 1.5$~m for the calculation of mixing length in Eq.~\eqref{eq:scalar}.

Four representative flow fields obtained from simpleFoam are displayed in Fig.~\ref{fig:flowfields}, showing the streamlines for the cases with limiting slopes in the training flows (Figs.~\ref{fig:flowfields}a and~\ref{fig:flowfields}b for $\alpha=1$ and 2, respectively) and testing flows (Figs.~\ref{fig:flowfields}c and~\ref{fig:flowfields}d for $\alpha=0.5$ and 4, respectively). The backgrounds of the plots highlight the recirculation zones with contours of the streamwise velocities $u_x$ with blue (darker grey) near the bottom wall showing the regions with flow reversal. The testing flows are intentionally chosen to span a wide range of flow patterns ranging from massive separation ($\alpha=0.5$ with the recirculation zone extending to windward of the hill) to mild separation ($\alpha=2$ with recirculation bubble at the leeward foot of the hill) and fully attached flow ($\alpha=4$). Such a wide range of flow patterns is expected to pose challenges to the closure modeling from flow field to concentration.

\begin{figure}[!htb]
\centering
\includegraphics[width=0.8\textwidth]{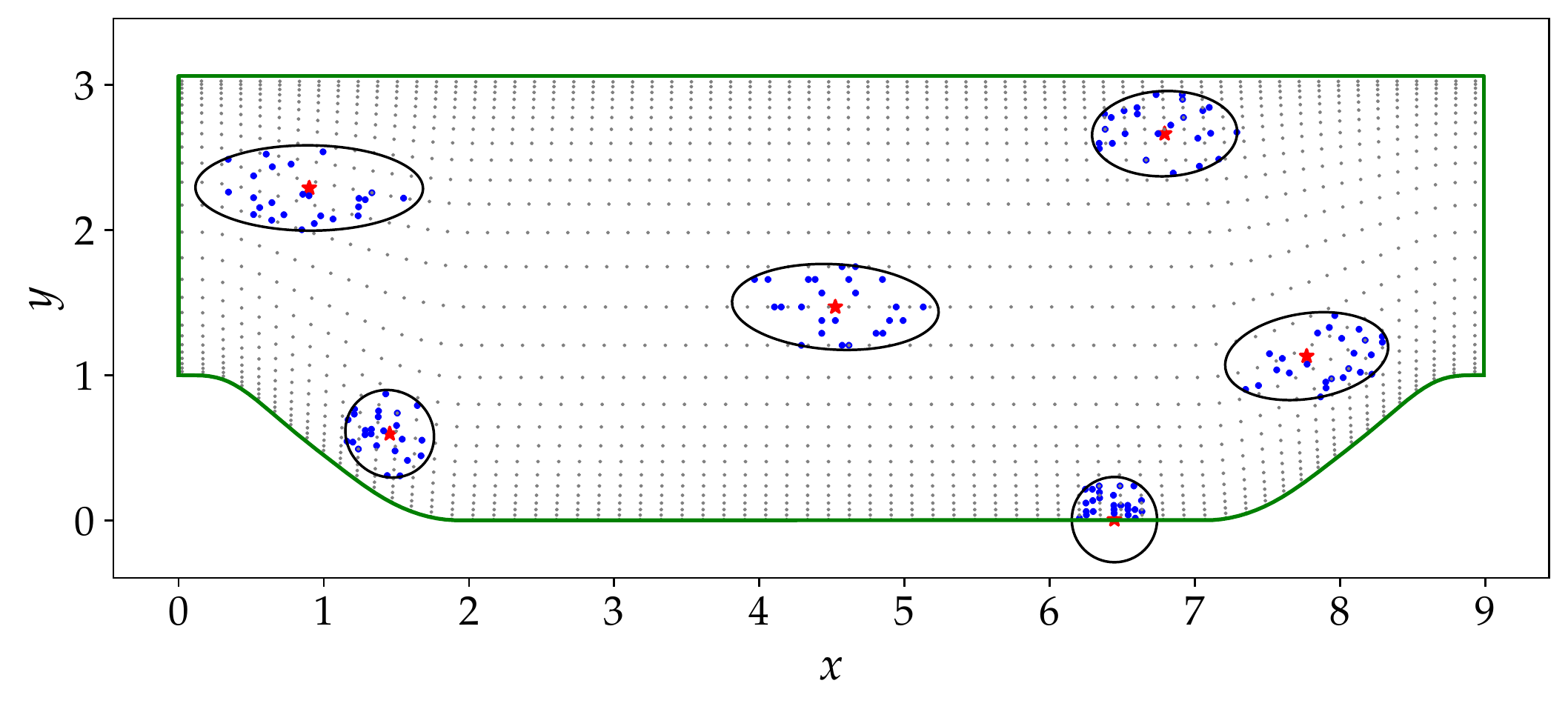}
  \caption{
  Method of sampling data points to generate labeled training data mapping a vector cloud $\cQ$ to the closure variable $\tau$. The gray dots ($\color{gray!80} \bullet$) indicate cell centers, showing only every seventh row and third column for clarity.  The surrounding cloud is depicted as  ellipses (\protect\tikz{ \protect\draw (0,-5) ellipse (6pt and 2.5pt);}), whose size and orientation are determined by the velocity at the cloud center ($\color{red} \star$) according to the region of physical influence. 
  The cloud centers ($\color{red} \star$) indicate the locations where the concentration $\tau$ is to be predicted.
  The blue/darker gray dots ($\color{blue} \bullet$) are randomly sampled from the cell centers ($\color{gray!80} \bullet$) within the cloud, and the feature vectors attached to them are used as input matrix $\cQ$ to predict~$\tau$.}
  \label{fig:stencil-view}
\end{figure}

We aim to learn a nonlocal mapping from a patch of flow field $\region{\mathbf{u}(\bm{x})}$ to the concentration $\tau$ at the point of interest. The procedure of generating such pairs of data ${(\cQ, \tau)}$ is illustrated in Fig.~\ref{fig:stencil-view}, where $\cQ$ is the input feature matrix derived from the patch of flow field $\region{\mathbf{u}(\bm{x})}$. In Fig.~\ref{fig:stencil-view}, the cell centers on which the flow fields are defined are indicated as grey dots ($\color{gray!80} \bullet$), showing only every seventh row and third column for clarity; 
the highlighted point ($\color{red} \star$) indicates the location where the concentration is predicted, and the ellipse (\tikz \draw (0,-5) ellipse (6pt and 2.5pt);) denotes the surrounding vector cloud; $n$ data points ($\color{blue} \bullet$) are randomly sampled within the cloud for the prediction of concentration $\tau$.
The extent of the cloud is determined by the velocity $\bu$ at the point of interest according to region of physical influence. The major axis of the ellipse aligns with the 
direction of the velocity $\bu$. The half-lengths $\ell_1$ and $\ell_2$ of the major and minor axis are determined based on the specified relative error tolerance $\epsilon$ according to~\cite{zhou2020nonlocal}:
\begin{equation}
    \ell_{1} = \bigg|\frac{2\nu\log\epsilon}{\sqrt{\vnorm{\bu}^2+4\nu\zeta}-\vnorm{\bu}}\bigg| \quad \text{and} \quad \ell_{2} = \bigg|\sqrt{\frac{\nu}{\zeta}}\log\epsilon\bigg|,
    \label{eq:l}
\end{equation}
where $\nu$ and $\zeta$ are the coefficients for diffusion and dissipation terms in the one-dimensional steady-state convection-diffusion-reaction equation $-\nu \frac{\mathrm{d}^{2}}{\mathrm{~d} x^{2}} \tau(x)+u \frac{\mathrm{d}}{\mathrm{d} x} \tau(x)+\zeta \tau(x)=P(x)$, and the detailed derivation can be found in~\cite{zhou2020nonlocal}. In principle, the actual dissipation coefficient $\zeta$ in Eq.~\eqref{eq:l} varies in space as $C_\zeta \tau$, considering the dissipation term $\mathsf{E} =  C_\zeta \tau^2$. Here, we assume it to be constant $\zeta \approx C_\zeta\tau_{\max}$ to control the size of clouds, in which $\tau_{\max}$ is obtained by taking the maximum value from the steady-state concentration field. The diffusion coefficient $\nu$ in Eq.~\eqref{eq:l} is chosen as the coefficient $C_\nu$ in Eq.~\eqref{eq:scalar}.

For convenience, we set $n$, the number of sampled data points in a cloud, to 150 in all training data, which gives good predictions in our problems.
It should be noted that, the constructed region-to-point mapping is applicable to data with different numbers of sampled data points within the cloud. We set it to be constant in the training data to process them in batch with better computation efficiency.
Given a fixed number of sampled data points within a cloud, the sampling method influences the performance of the trained model. Although the points closer to the center usually have a larger impact on the center due to the underlying convection-diffusion process, all the points in the elliptical region may have non-negligible contributions to the final prediction.
Always choosing the closest data points will indirectly reduce the size of the influence region, which may lose the effective nonlocal mean flow information and undermine the prediction performance~\cite{zhou2020nonlocal}.
Instead, we use the uniform random sampling method to sample $n$ data points from the cloud. In other words, we treat each data point equally a priori and let the embedding network learn the importance of each point implicitly.
In testing, we showcase the flexibility of our trained networks to the input data with different numbers of sampled points in the cloud by using all the data points within the cloud, whose number varies at different locations, to predict the concentration at the points of interest. 
In total, we generated $1.26 \times 10^6$ pairs of ${(\cR, \tau)}$ as training data, which takes around 6 hours on a single CPU. Note that we can use fewer pairs for training since the adjacent data pairs from the same geometry appear similar. In this study, we choose all the pairs to make full use of the generated data.
The proposed neural network is implemented and trained by using the machine learning library PyTorch~\cite{paszke2019pytorch}, which takes around 16.7 hours on an NVIDIA GeForce RTX 3090 GPU. We verified that the constructed constitutive neural network does ensure all the expected (translational, rotational, and permutational) invariance. The detailed architecture for the embedding and fitting networks as well as the parameters used  for training are presented in~\ref{app:nn}.
The code developed and data generated for this work are made publicly available on GitHub~\cite{zhou2021vcnn-git}. 

\begin{figure}[!htb]
\centering
{\includegraphics[width=0.45\textwidth]{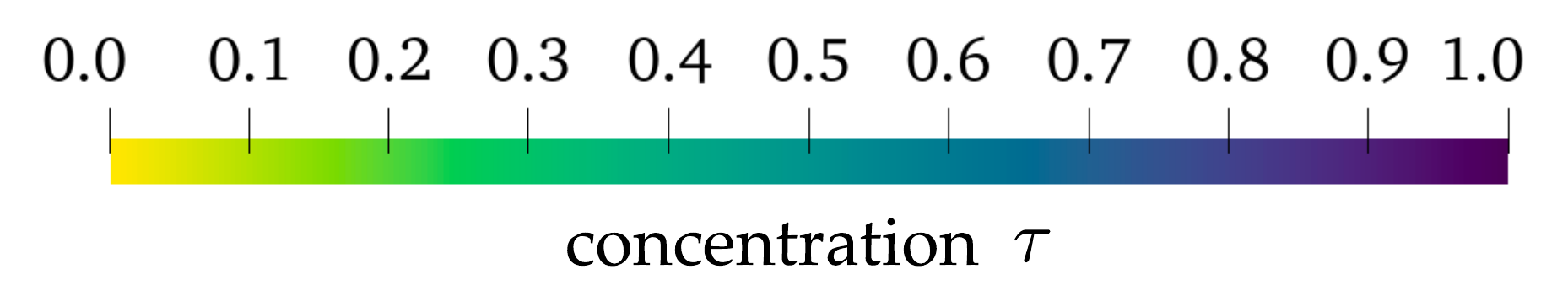}
\vspace{-8pt}}\\
\subfloat[ground truth, $\alpha = 0.5$]
{\includegraphics[height=0.10\textwidth]{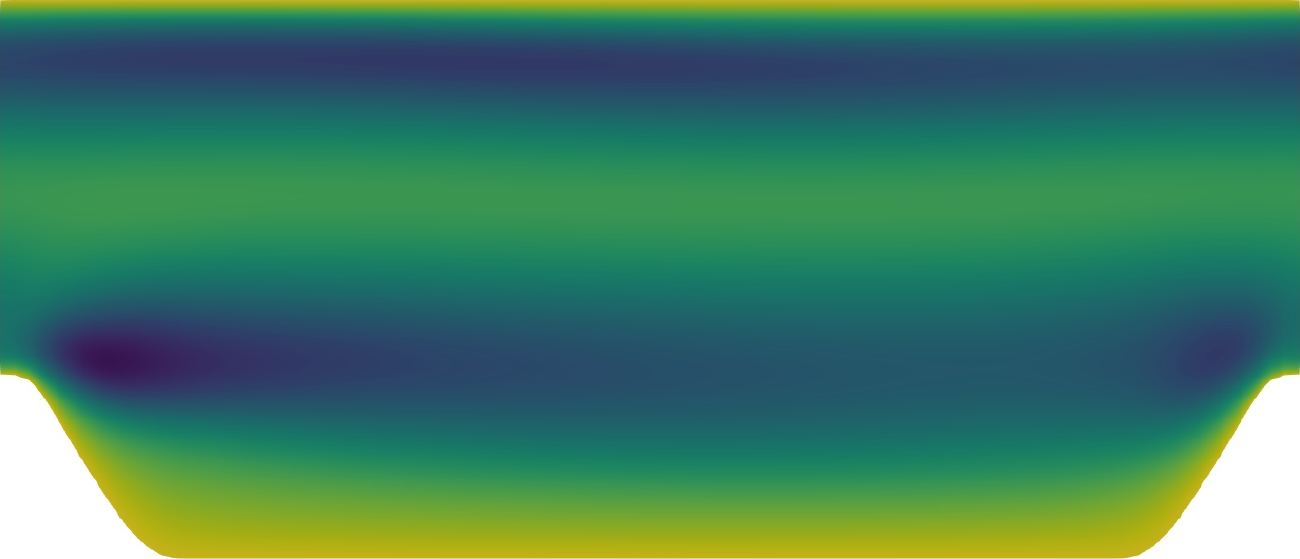}}
\hspace{0.1em}
\subfloat[ground truth, $\alpha = 4$]
{\includegraphics[height=0.10\textwidth]{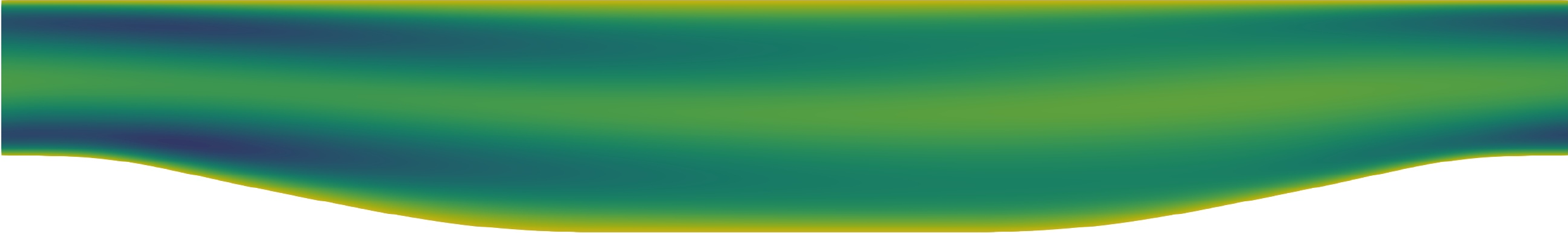}}\\
\subfloat[prediction, $\alpha = 0.5$]
{\includegraphics[height=0.10\textwidth]{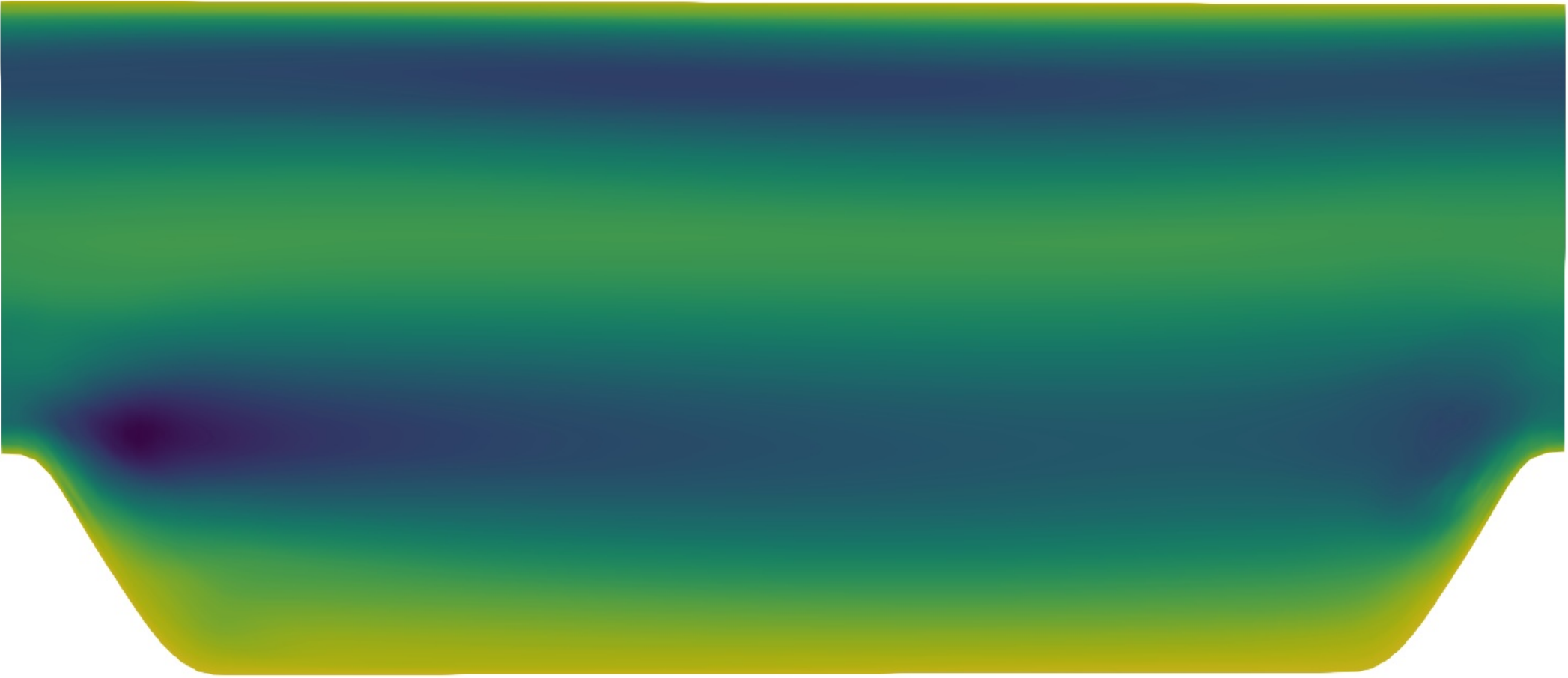}}
\hspace{0.1em}
\subfloat[prediction, $\alpha = 4$]
{\includegraphics[height=0.10\textwidth]{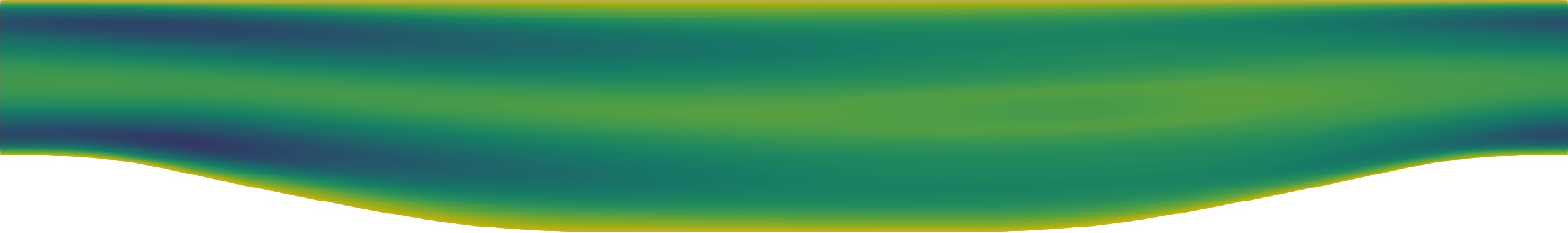}}\\
{\vspace{5pt}
\includegraphics[width=0.34\textwidth]{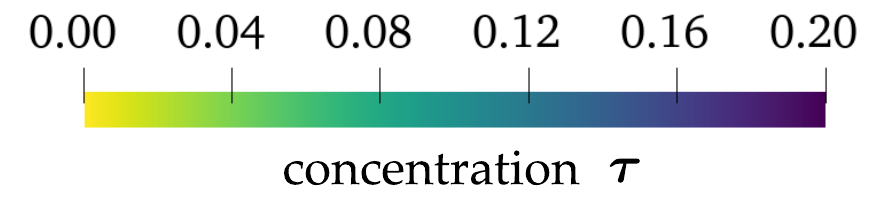}
\vspace{-10pt}}\\
\subfloat[difference, $\alpha = 0.5$]
{\includegraphics[height=0.103\textwidth]{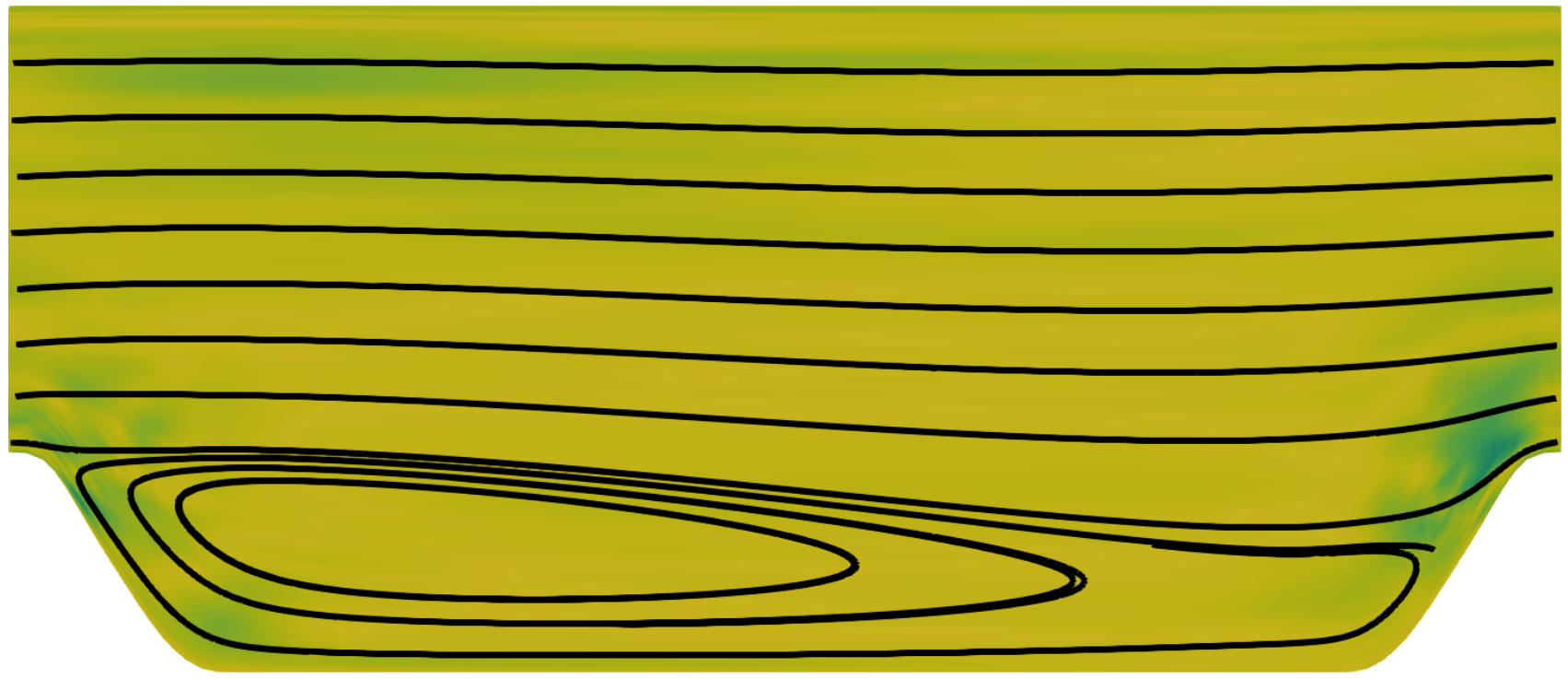}}
\hspace{0.1em}
\subfloat[difference, $\alpha = 4$]
{\includegraphics[height=0.103\textwidth]{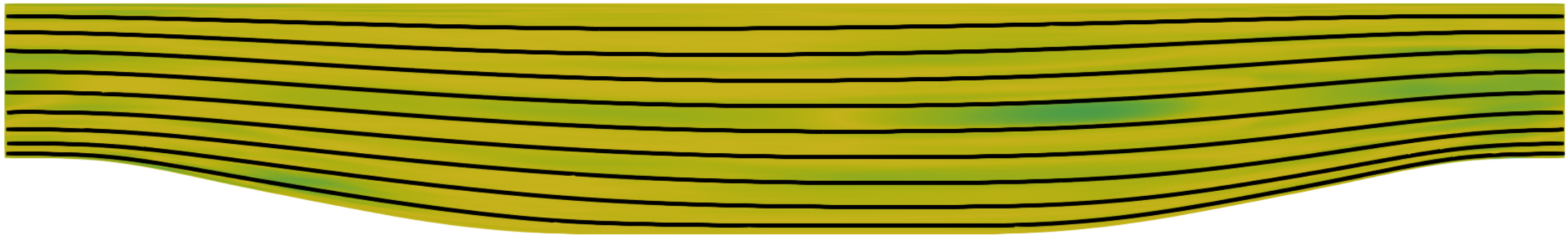}}\\
{\vspace{5pt}
\includegraphics[width=0.38\textwidth]{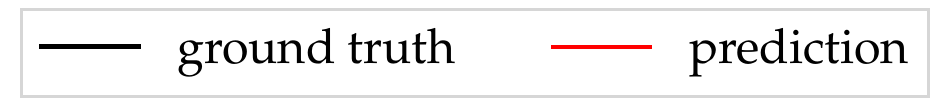}
\vspace{-10pt}}\\
\subfloat[cross-section, $\alpha = 0.5$]
{\includegraphics[width=0.27\textwidth]{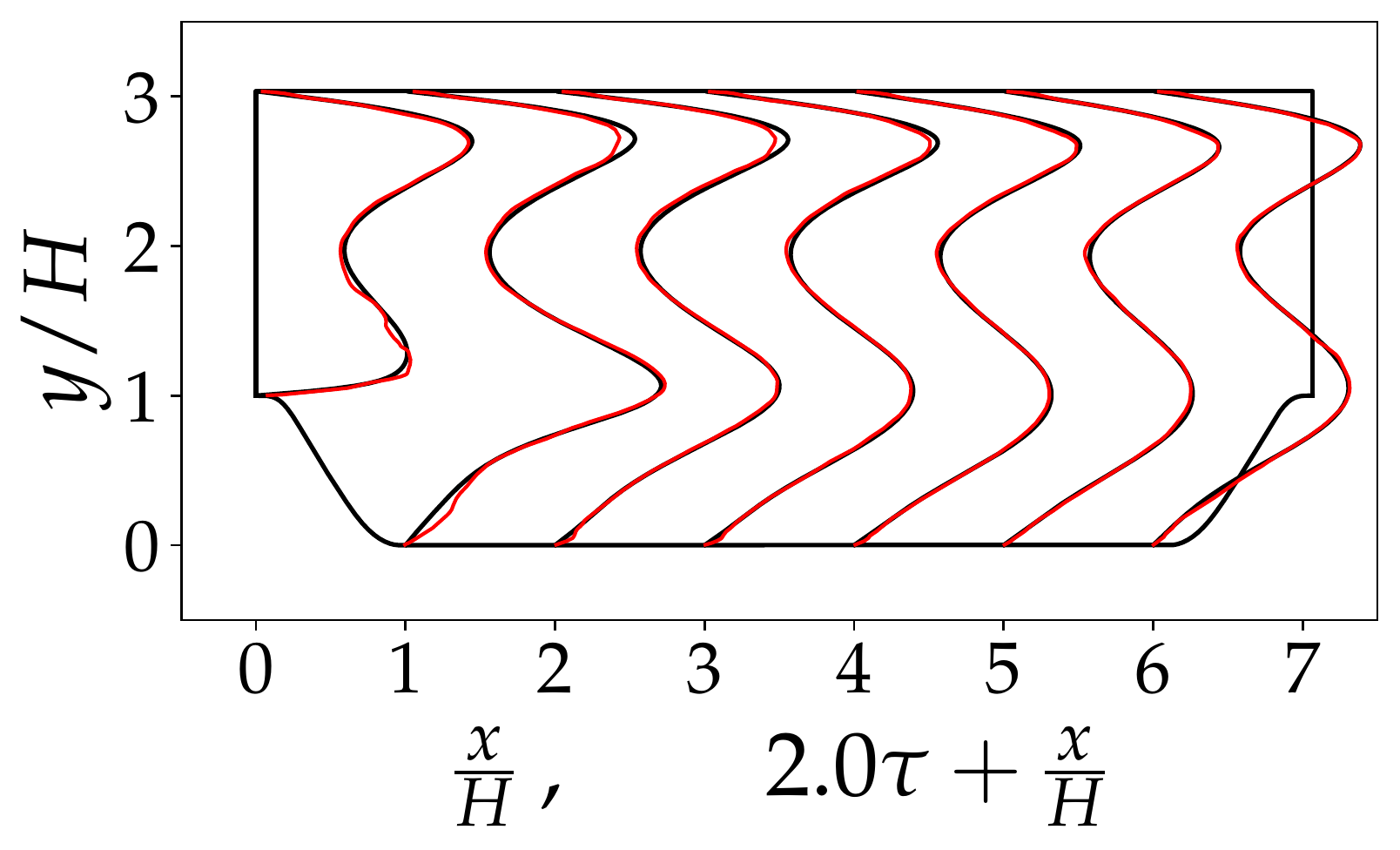}}
\hspace{0.1em}
\subfloat[cross-section, $\alpha = 4$]
{\includegraphics[width=0.67\textwidth]{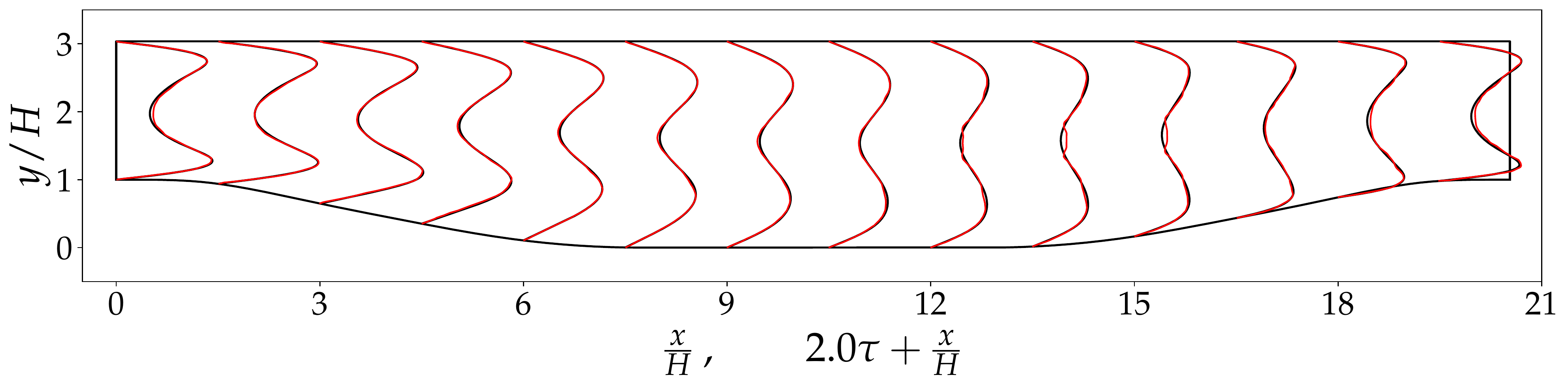}}
  \caption{Comparison of the ground truths of the concentration field $\tau$ (top row), the corresponding predictions with the trained constitutive neural network (second row) and the differences between the ground truths and the predictions (third row) along with the concentration plots in seven and fourteen cross-sections (bottom row) for two configurations with slope parameters $\alpha=0.5$ (left panels) and $\alpha =4.0$ (right panels). The neural network is trained with data from 21 configurations with $\alpha = 1, 1.05, 1.1, \ldots, 2$.
  }
  \label{fig:concentration-compare}
\end{figure}

\subsection{Neural-network-based prediction of concentrations}
After training, the predicted concentration fields provided by the trained constitutive network are quite close to the ground truths. The comparison of the predicted concentration fields and the corresponding ground truths in two extreme configurations with $\alpha =$ 0.5 and 4 is shown in Fig.~\ref{fig:concentration-compare}.
The two flow fields have distinctive patterns in that there is a large recirculation zone for $\alpha =0.5$ as shown in Fig.~\ref{fig:flowfields}.
According to Fig.~\ref{fig:concentration-compare}a--\ref{fig:concentration-compare}e, we can see that although neither field is present in the training data, both predicted concentration fields are similar to the corresponding ground truths.
The accuracy is visualized more clearly in Fig.~\ref{fig:concentration-compare}g, the comparison of the predicted and true concentration on seven vertical cross-sections at $x/H = 0, 1, \cdots, 6$ when $\alpha = 0.5$. The same comparison is presented in Fig.~\ref{fig:concentration-compare}h for the case $\alpha = 4$ on fourteen cross-sections.
From Fig.~\ref{fig:concentration-compare}e and Fig.~\ref{fig:concentration-compare}g, we can see that the predicted concentrations near both the inlet and outlet show slight inconsistency with the ground truths in the case $\alpha = 0.5$, which may be caused by the flow separation due to the sharp hill. According to Fig.~\ref{fig:concentration-compare}f and Fig.~\ref{fig:concentration-compare}h, however, the inconsistency is not concentrated near the hills in the case $\alpha=4$ since the slope is very gentle. The slight inconsistency in the middle channel near $x/H \approx 15$ is mainly caused by the difference from the training flows, i.e., the main stream here is being contracted along a longer hill while it is less contracted in the training flows due to the shorter hills and the existence of the recirculation zones.

\begin{figure}[!htb]
\centering
{\includegraphics[width=0.99\textwidth]{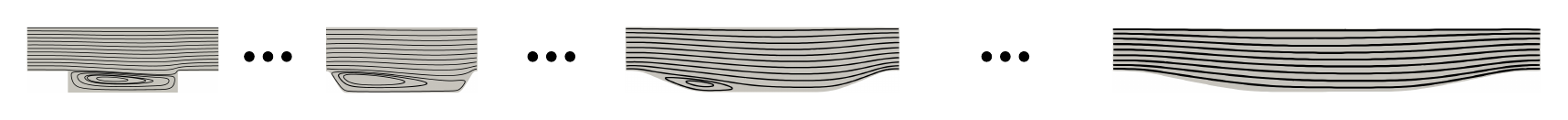}
\vspace{-8pt}}\\
\includegraphics[width=0.99\textwidth]{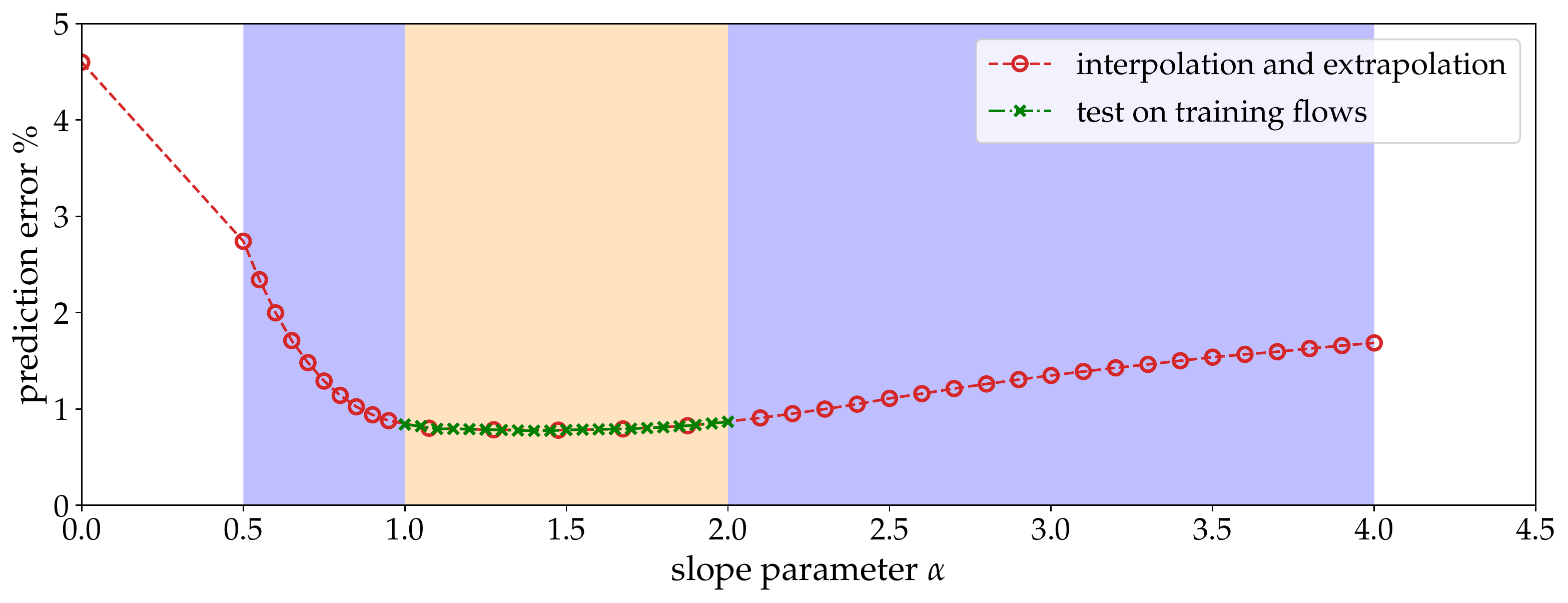}
  \caption{
  General predictive errors at various slope parameters $\alpha$. The yellow/lighter and blue/darker backgrounds represent the regimes of the slope parameters $\alpha$ of training and testing flows, respectively. The neural network is trained with data from 21 configurations with $\alpha = 1, 1.05, 1.1, \ldots, 2$.
  The trained network is then tested on five configurations with $\alpha = 1.075, 1.275, \ldots, 1.875$ in the training regime and 30 configurations with $\alpha = 0.5, 0.55, \ldots, 0.95$ and $\alpha = 2.1, 2.2, \ldots, 4$ in the testing regime. The thumbnails on the top indicate the variation of geometry and flow field with slope parameter $\alpha$ increasing from 0.5 to 4. We also tested a configuration with a sudden expansion and contraction, roughly corresponding to $\alpha = 0$.}
  \label{fig:general-testing}
\end{figure}

We use 35 configurations with varying slope parameters $\alpha$ to evaluate the prediction performance of the trained neural network. The prediction errors of testing flows are shown in Fig.~\ref{fig:general-testing}. The normalized prediction error is defined as the normalized pointwise discrepancy between the predicted concentration $\hat{\tau}$ and the corresponding ground truth $\tau^*$:
\begin{equation}
    \text{error} = \frac{\sqrt{\sum_{i=1}^{N} {|\hat{\tau}_i -
    \tau_{i}^*|}^2}}{\sqrt{\sum_{i=1}^{N} {|\tau_{i}^*|}^2}}, 
    \label{eq:error-def}
\end{equation}
where the summation is performed over all of the $N$ training or testing data points. The in-between (yellow/lighter gray) region and the whole region represent the regimes of slope parameter $\alpha$ of training flows and testing flows, respectively. We can see that the prediction errors of interpolated testing flows with $\alpha = 1.075, 1.275, \ldots, 1.875$ are the lowest, which is expected because the flow fields in these configurations are very close to that of the training flows. For extrapolated testing flows, the prediction error grows as $\alpha$ departs from the range 1--2 in the training data (yellow/lighter gray). Specifically, as $\alpha$ approaches 4 the recirculation zone gradually shrinks and eventually disappears, and the prediction error grows. In contrast, when $\alpha$ decreases to 0.5, the whole bottom region becomes the recirculation zone, and the prediction error increases relatively fast due to the massive flow separation that is dramatically different from those in the training flows. Further, we tested the trained network on a geometry with a sudden expansion at $x/H = 1.93$ and contraction at $x/H = 7.07$ (roughly corresponding to $\alpha = 0$). The prediction error was 4.76\%, which was higher than that for $\alpha = 0.5$ but still reasonable.
In addition, we also performed testing on 21 training flows. 
Note that in this testing step all the data points in a cloud (approximately 300 to 1500 points) are used for prediction, although only 150 points within each cloud were used in training. The prediction errors on 21 training flows are all around 1\%, which is close to the training error 0.62\%.
This shows that the proposed nonlocal constitutive neural network not only guarantees permutational invariance \emph{formally}, but also approximates the underlying PDE mapping with adequate accuracy and great flexibility in terms of the number and locations of the used points.

\begin{figure}[!htb]
\centering
\includegraphics[width=0.95\textwidth]{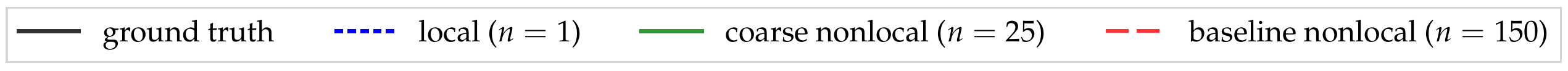}\\
\includegraphics[width=0.95\textwidth]{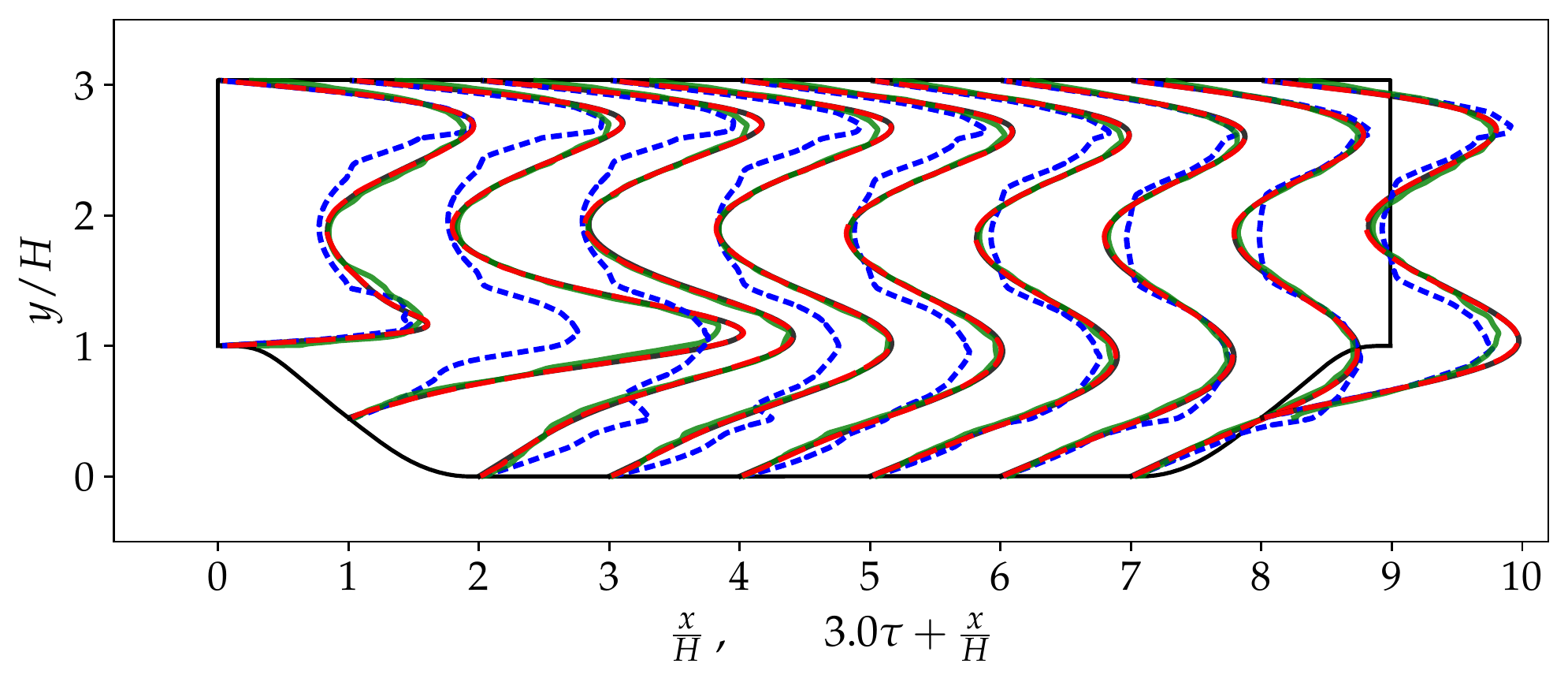}
  \caption{Ablation study results showing a comparison of the predicted concentrations on nine cross-sections in the flow with $\alpha =1$ by using different models, including a local model ($n=1$; blue dotted lines) and nonlocal models with a coarse sampling ($n=25$ points per cloud; green/light grey solid lines) and a baseline sampling ($n=150$ points; red dashed lines). The results from $n=150$ are visually indistinguishable from the ground truth.  The neural networks are trained with data from 21 configurations with $\alpha = 1, 1.05, 1.1, \ldots, 2$.
  }
  \label{fig:cloud-size-compare}
\end{figure}

While we use $n=150$ points per cloud to achieve optimal results, our ablation studies show that (1) using a nonlocal model with coarser cloud sampling causes deteriorated performance due to the sampling error, and (2) using a local model would diminish the predictive capability of the model as it completely misses the nonlocal physics in the data.
To this end, we investigate two levels of ablated models: a coarse nonlocal model and a local model. First, we study a coarse nonlocal model by randomly choosing $n=25$ data points within the clouds and train the neural network with such generated data. The trained neural network is then tested based on all the data points within the clouds. In a further ablation, we evaluate a local model with $n=1$ point, i.e., only the data point at the center is used for training (without information from any neighboring points). In the latter case, the local constitutive model as represented by the neural network no longer has pairwise inner productions. As a result, only the scalar quantities $\bw = [\theta, s, b, \mathsf{u}, \eta]^\top$ attached to the center point are used for training and they are directly fed into the fitting network (Fig.~\ref{fig:framework}c). An embedding network (Fig.~\ref{fig:framework}b) is thus omitted in the local model. Accordingly, the trained neural network is tested by feeding the scalar quantities $\bw$ associated with the data point itself and not the cloud. The comparison of the three different models is shown in Fig.~\ref{fig:cloud-size-compare}. 
It can be seen that the predicted profiles from the local model ($n=1$, blue/dark grey dotted lines) deviate significantly from the ground truth in most  regions. In contrast, the prediction of nonlocal models (green and red dashed lines) are almost indistinguishable from the ground truth, although the sparsely sampled model (with $n=25$ points per cloud) deviates from the truth in a few places.
For a quantitative comparison, the local model does not perform well and the prediction error rate is much larger (19.2\%) even when compared to the error rate 7.65\% of the sparsely sampled nonlocal model. This is expected as the nonlocal model completely misses the nonlocal physics due to convection and diffusion. The prediction error is further reduced to 0.84\% when a denser sampling $n=150$ is used in the nonlocal model.

\subsection{Physical interpretation of learned model}

The learned network implies a weight (contribution) for each point in any given cloud, and it is desirable to shed light on such a weight and its dependence on the location of the point in the cloud. Physical intuition dictates that the weight shall be skewed towards in the upwind direction, and that such a skewness shall increase (diminish) as the local velocity increases (diminishes).
To verify such a trend, we chose the embedded basis $\cG^\star \in \bbR^{n \times m'}$ as a proxy of the weight for each point. Furthermore, we set $m'=1$ so that there is exactly one number associated with each of the $n$ points. We chose five points in three representative regions for visualization: (i) a point $\text{M}$ in the mid-channel, (ii) two points R-$135^\circ$ and R-$180^\circ$ in the recirculation zone (named according to their respective velocity angles in reference to the streamwise direction), (iii) two points $\text{W}_1$ and $\text{W}_2$ near the top and bottom walls, respectively. 

\begin{figure}[!htb]
\centering
\subfloat[location, cloud, and local velocity]
{\includegraphics[height=0.29\textwidth]{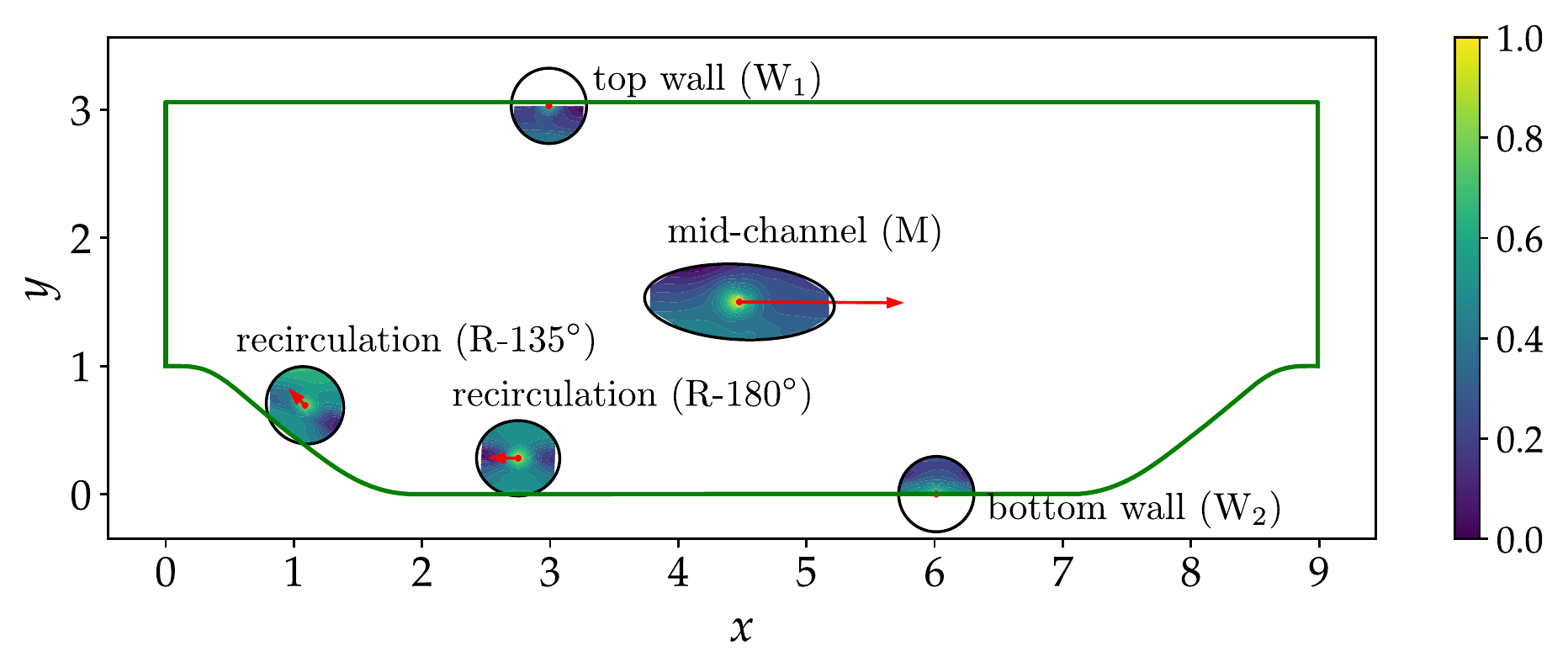}}
\hspace{1em}
\subfloat[weight contours]
{\includegraphics[height=0.29\textwidth]{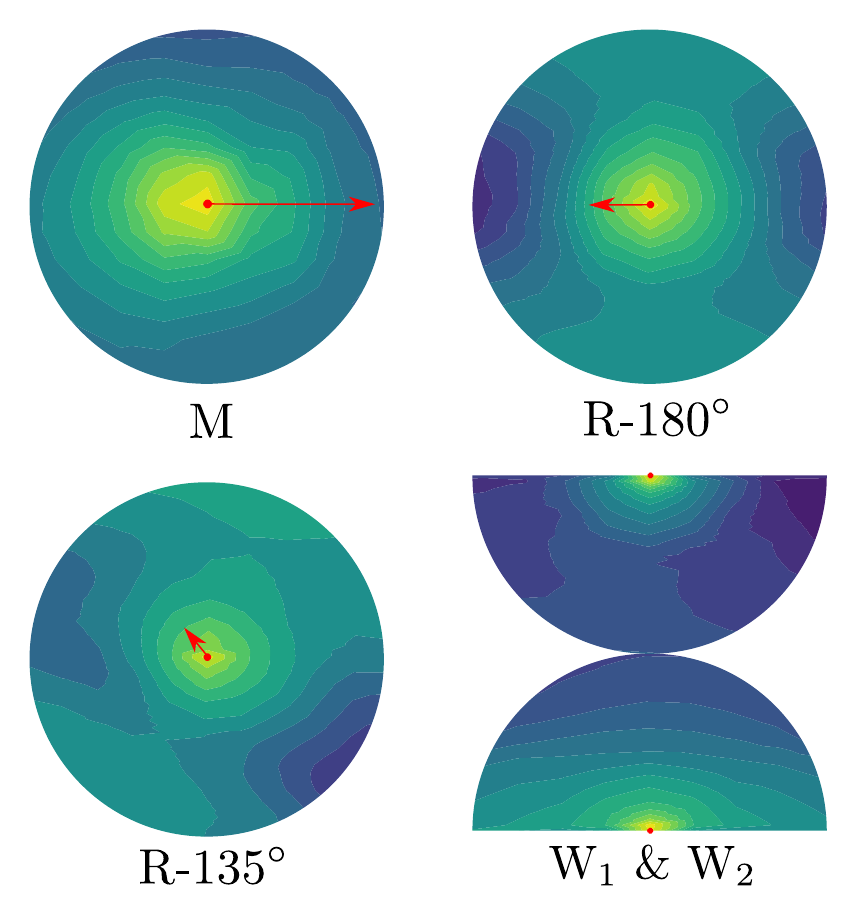}}\\
{\includegraphics[width=0.5\textwidth]{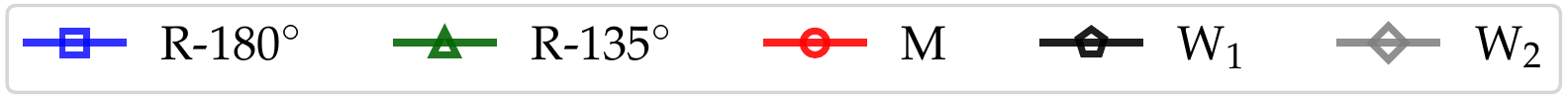}
\vspace{-10pt}}\\
\subfloat[profiles along local velocity]
{\includegraphics[height=0.31\textwidth]{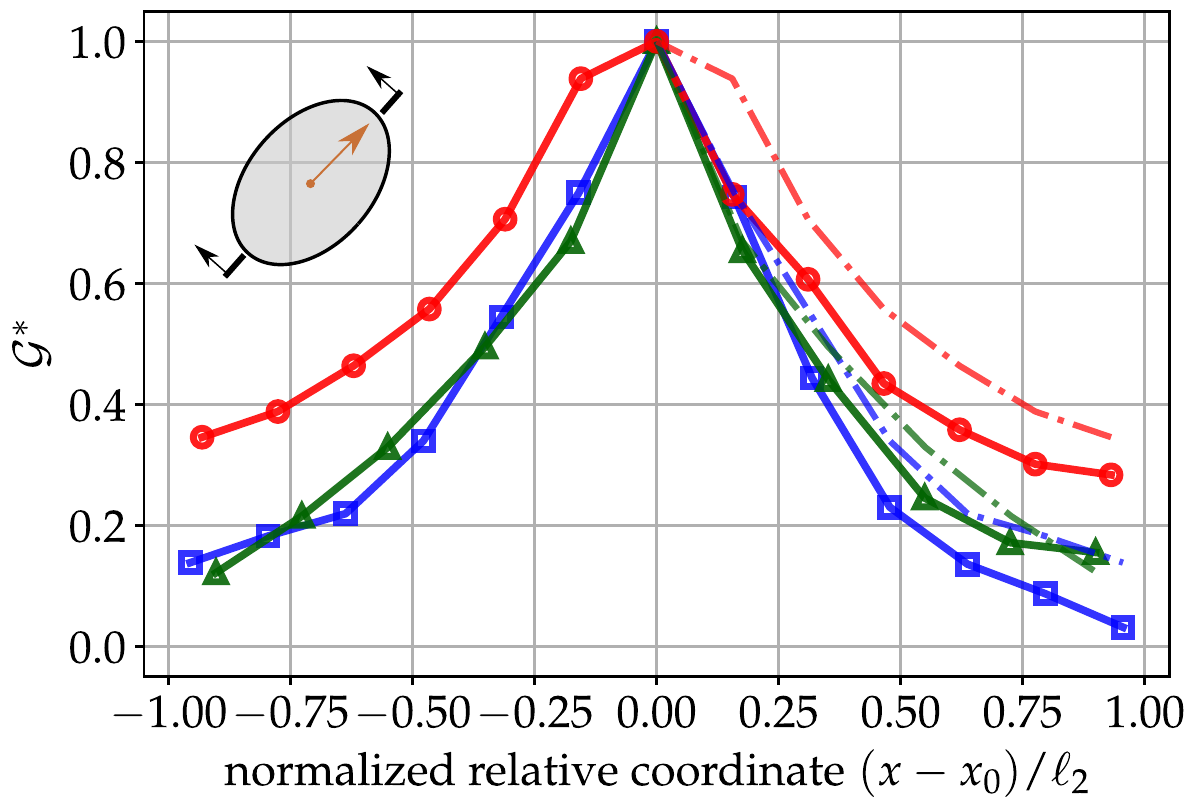}}
\hspace{1em}
\subfloat[profiles perpendicular to local velocity]
{\includegraphics[height=0.31\textwidth]{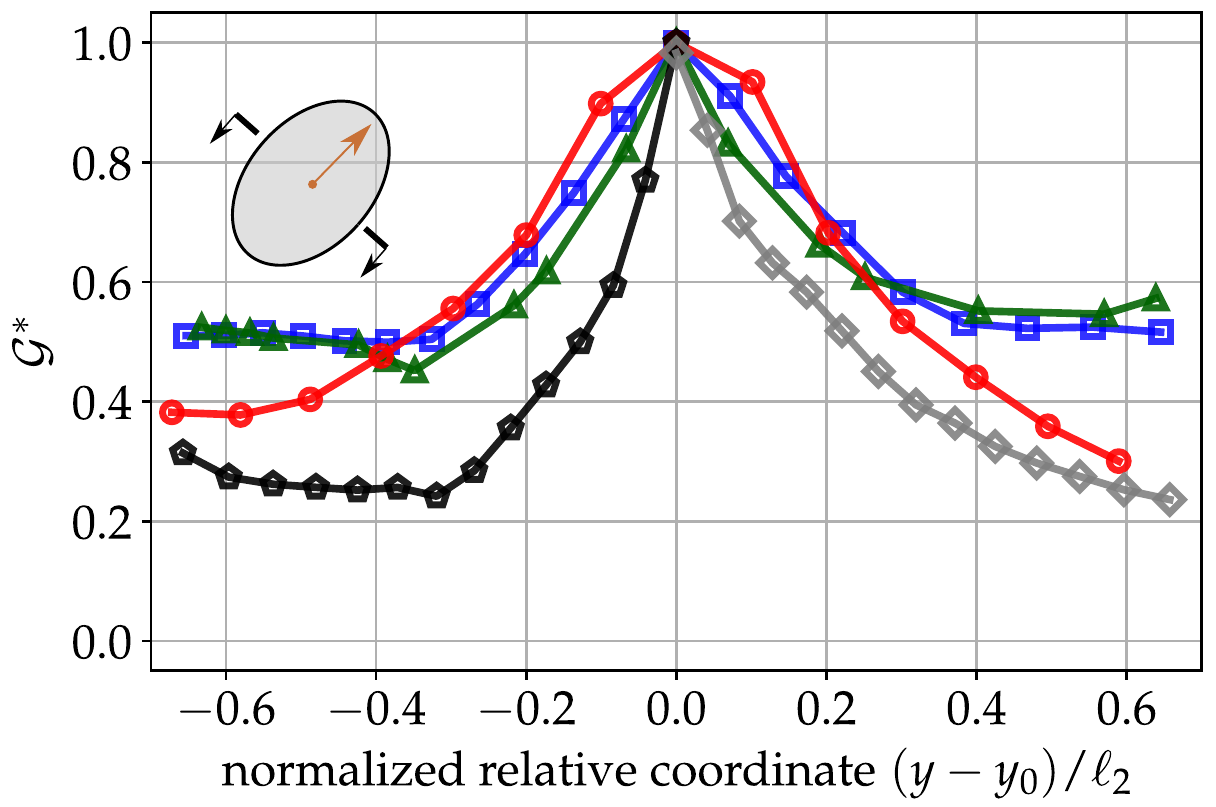}}\\
  \caption{
  Weight $\cG^\star$ for each point in the cloud as learned by the neural network in the case of $m'=1$, showing skewness towards the upwind direction as expected from physical intuition. Weight contours are shown for five points in three representative regions: M in the mid-channel, R-$135^\circ$ and R-$180^\circ$ in the recirculation zone, and $\text{W}_1$ and $\text{W}_2$ near the walls. (a) Locations, clouds (indicated in ovals), and local velocities (indicated in red/grey arrows) for each point; (b) Weight contours zoomed to the circle of radius~$\ell_2$; (c) Profiles along the local velocity direction, with the upstream mirrored to downstream as dash-dotted lines; (d) Profiles perpendicular to local velocity direction. Note that the weight $\mathcal{G}^{\star}$ is normalized by the maximum $\mathcal{G}^{\star}$ in the cloud such that the plotted weight has a maximum of 1.
  }
  \label{fig:G-analysis}
\end{figure}

The contours of the weights showed a trend that is consistent with the physical intuition above. The locations, clouds and local velocities of these points are shown in Fig.~\ref{fig:G-analysis}a. The weight contours (showing only a circle of radius $\ell_2$ around the point) are presented in Fig.~\ref{fig:G-analysis}b, where the weights in the upwind regions are clearly larger for the cloud of the mid-channel point M, while for other points it is less prominent due to their smaller velocities. To show the trend more clearly, we present two cross-sections in Fig.~\ref{fig:G-analysis}c (along the local velocity) and Fig.~\ref{fig:G-analysis}d (perpendicular to the local velocity direction). In Fig.~\ref{fig:G-analysis}c, the upwind region (negative relative coordinates $x-x_0$) is mirrored to the downwind region and denoted in dash-dotted lines of the same color. This comparison clearly shows that the upwind region has larger weights. Our parametric studies further showed that such a trend of upwind skewness was even more evident when the diffusion coefficient $C_\nu$ was reduced when generating the training data (figures omitted here for brevity). In contrast, the skewness is not present in the profiles perpendicular to the local velocity as shown in Fig.~\ref{fig:G-analysis}d, which is expected since there is no convection in this direction. 
The general trend of the weights for the near wall points $\text{W}_1$ and $\text{W}_2$
are similar to those for the in-stream points (M, R-$135^\circ$, and R-$180^\circ$) because they must emulate the same differential operator. 
On the other hand, the weights for the near wall points $\text{W}_1$ and $\text{W}_2$ decrease more rapidly from center to periphery than those for the in-stream points (M, R-$135^\circ$, and R-$180^\circ$), meaning that the learned weights for points near the boundaries are significantly larger than those far from the boundaries. Such a difference in the trend is desirable: the embedding network needs to also emulate a wall-function implicitly embedded in the mixing length model in Eq.~\eqref{eq:ellm} for the near-wall points.
In summary, the neural network correctly learned from the data the contribution of each point in the cloud.

\subsection{Discussion}
\begin{figure}[!htb]
\centering
{\includegraphics[width=0.5\textwidth]{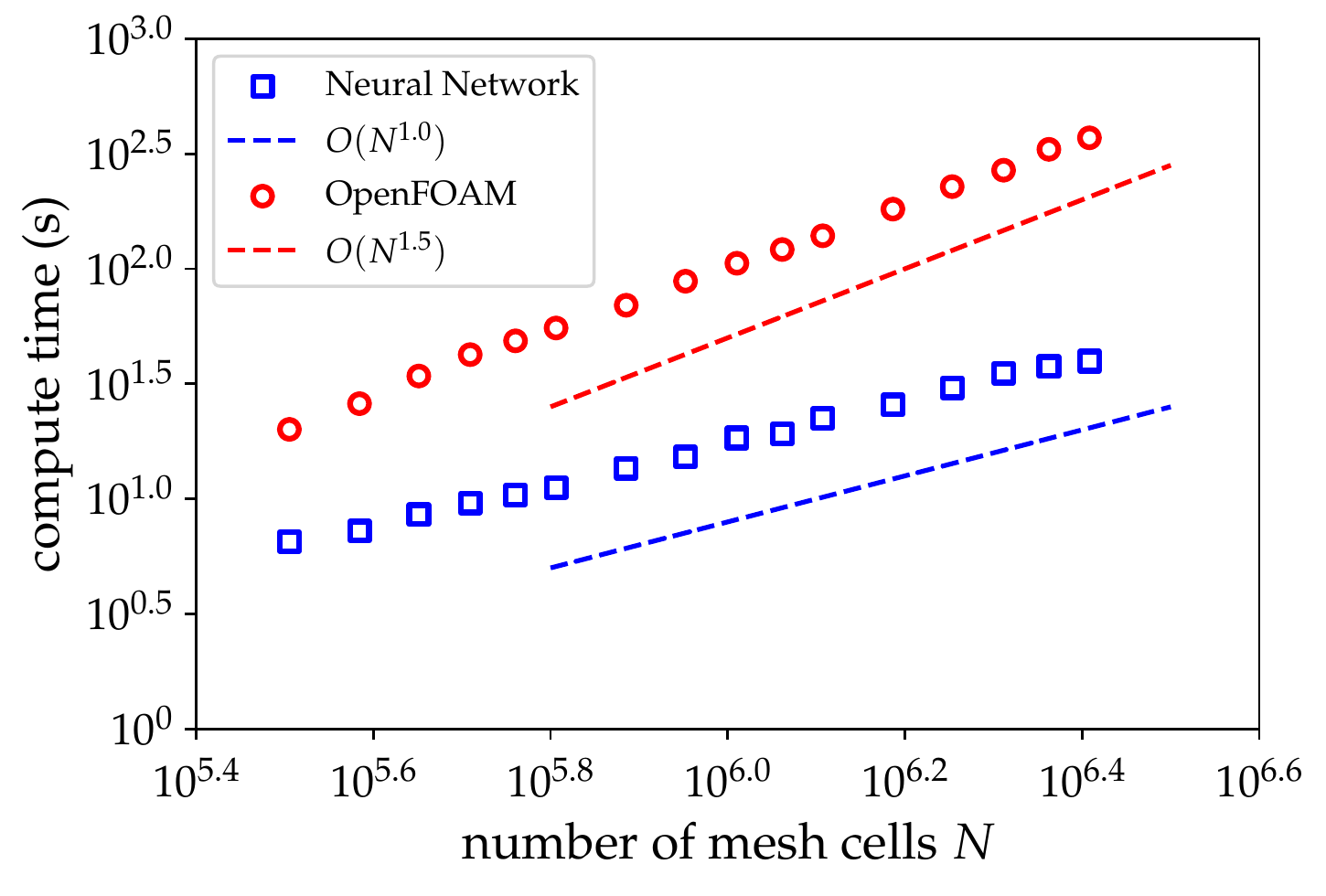}}
  \caption{Comparison of computational time and its scaling with problem size (number of cells $N$ in the mesh) between vector-cloud neural network and traditional PDE solver (finite volume method in OpenFOAM). Different cases correspond to periodic hills with the slope parameter $\alpha$ ranging from 0.5 to 4.}
  \label{fig:time-comparison}
\end{figure}

Besides the physical interpretability, by design the proposed vector-cloud neural network has some other merits that make it suitable for nonlocal closure modeling. On the training side, it is flexible with a wide range of data, which facilitates the process of data collection. Training samples are well defined if we only have data in a region rather than the whole field or the spatial meshes are different in different regions. On the inference side, the computational cost of predicting the tracer concentration of the entire field is proportional to the total number of mesh cells $N$, given a fixed number of sampled points in the cloud. Due to the lack of well-established machine-learning-based models for the closure modeling, we compare the computational complexity of the vector-cloud neural network with traditional PDE solver (finite volume method in OpenFOAM) on the periodic hills with the slope parameter $\alpha$ ranging from 0.5 to 4. The results are shown in Fig.~\ref{fig:time-comparison}. The linear scaling $O(N)$ of the vector-cloud neural networks is confirmed. In contrast, although the scaling of the traditional finite volume method inevitably depends on the specific linear solver used (e.g., direct solver, conjugate gradient, or multigrid method), it can never achieve linear scaling. In our test, the scaling of the OpenFOAM solver is approximately $O(N^{1.5})$.  Such scaling, however, is never constant. Solving constitutive PDEs can be accelerated when coupled with RANS equations since the constitutive PDEs have better initial conditions from the last iteration. But for real industrial flows, solving the PDEs in constitutive models is very likely to be expensive and time-consuming while using the neural network based constitutive models can be much faster. Note that this experiment is performed on a single CPU, but in practice the proposed network can be massively parallel since the computation of each point is independent, which would further increase the computational superiority of the neural-network-based method.

\section{Conclusion}
\label{sec:conlusion}
Constitutive closure models based on transport partial differential equations (PDEs) are commonly used in computational fluid dynamics. Such closure models often lack robustness and are too rigid to calibrate with diverse training data. Drawing inspirations from a previous neural-network-based molecular potential energy model, in this work we propose a vector-cloud neural network to represent the nonlocal physics in the transport PDEs as a region-to-point mapping. Such a vector-cloud network is invariant to coordinate translation, rotation, and ordering of the points. It also accounts for the transport physics through corresponding scalar features.
The vector-cloud network works on arbitrary grids and is suitable for finite-element and finite-volume solvers that are common in fluid dynamics. The proposed network has been used as surrogate models for scalar transport PDEs on a family of parameterized periodic hill geometries and it has been demonstrated to have desired invariance and  satisfactory accuracy. Therefore, it is a promising tool not only as nonlocal constitutive models and but also as general surrogate models for PDEs on irregular domains.

Despite the preliminary successes demonstrated in the present work, there are several directions of future research that should be conducted.
First and foremost, the performance of using the learned nonlocal constitutive models as the closure models for solving the primary PDEs (e.g., the RANS equations in turbulence modeling for mean velocities and pressure) should be methodically examined in future work.
Moreover, the proposed neural network model is now applied to a passive scalar, still different from the full Reynolds stress tensor in the RANS equations. Learning a constitutive model for a general anisotropic tensor is not straightforward. We need to guarantee the rotational equivariance of the tensor instead of the rotational invariance of the scalar. The work of predicting the polarizability tensor of liquid water~\cite{sommers2020raman} based on Deep Potential provides a possible approach to achieve this goal.
Nevertheless, we still need to explore more options to accurately model the complicated dynamics of the Reynolds stress in turbulence with neural networks.
Last but not least, we have demonstrated the invariance of the proposed vector cloud neural network but not shown the advantages of such invariance, especially the rotational invariance. The superiority should be studied systematically by comparing with the neural network without such invariance in future work.

\appendix
\section{Construction of constitutive scalar transport equation}
\label{app:tke}
The transport PDE in Eq.~\eqref{eq:scalar} is used in this work as target for the vector-cloud neural network to emulate. Its construction is inspired by the transport PDEs encountered in turbulence modeling. Among the most commonly used turbulence models are the eddy viscosity models. The transport of turbulent kinetic energy (TKE) $k$ for incompressible flows is described by the following PDE~\cite{pope00turbulent}:
\begin{equation}
 \bu \cdot \nabla k - \nabla \cdot \big((\nu+\nu_\text{t}) \nabla k \big)  = \mathsf{P} - \mathsf{E} \notag
\end{equation}
where $k = \frac{1}{2}\operatorname{tr}(\bm{\tau})$ is the trace of the Reynolds stress tensor $\bm{\tau}$, $\mathbf{u}$ is the mean (Reynolds-averaged) velocity, $\nu_\text{t}$ is the turbulent eddy viscosity, $\mathsf{P} = \bm{\tau} : \nabla \mathbf{u}$ is the production term, and $\mathsf{E}$ is the dissipation term. 

In order to provide closure for $\bm{\tau}$ in the production term, we will utilize the Boussinesq assumption $\bm{\tau} = 2 \nu_t \bm{s} - \frac{2}{3}k I$ along with the following facts:
\begin{equation}
    \nabla \mathbf{u} = \bm{s} + \Omega,  \quad
    \bm{s}: \Omega = 0, \quad  \text{and} \quad
    I : \nabla \mathbf{u} = \delta_{ij} \frac{\partial u_i}{\partial x_j} = \frac{\partial u_i}{\partial x_i} = 0 , \notag
\end{equation}
where $\bm{s}$ and $\Omega$ are strain-rate and vorticity tensors, respectively, $I$ (or Kronecker delta $\delta_{ij}$) is second-rank identity tensor, $:$ indicates double dot product (tensor contraction).  The production of TKE can thus be written as:
\begin{equation}
    \mathsf{P} = \bm{\tau} : \nabla \mathbf{u} 
    = 2 \nu_t \bm{s} : \nabla \mathbf{u}
    = 2 \nu_t \bm{s} : (\bm{s} + \Omega) = 2 \nu_t \, \bm{s}:\bm{s} \equiv \sqrt{k} \ell_m s^2
    \notag
\end{equation}
where in the last step the tensor norm is defined as $s = \|\bm{s}\| = \sqrt{2 \bm{s}:\bm{s}}$, and the turbulent eddy viscosity is approximated as $\nu_t = \sqrt{k} \ell_m$  based on Prandtl's mixing length assumption, with mixing length $\ell_m$ defined in Eq.~\eqref{eq:ellm}.
Evoking again the analogy between concentration $\tau$ and the TKE $k$, the boundary condition for $\tau$ is set to zero at the walls.
The dissipation of TKE in Prantdl's one-equation model follows the form $\mathsf{E} = C_\zeta \frac{k^\frac{3}{2}}{\ell_m}$. However, for simplicity, we set $\mathsf{E} = C_\zeta k^2$ (where the coefficient $C_\zeta$ shall be chosen to make the dimension consistent) by following the other transport equations~\cite{pope00turbulent}, e.g., the dissipation $\varepsilon$, the dissipation rate $\omega$, and the turbulent viscosity-like variable $\tilde{\nu}$, where the destruction terms are as $\varepsilon^2$, $\omega^2$, and $\tilde{\nu}^2$ in the respective equations. With $\mathsf{P} = \sqrt{k} \ell_m s^2$ and $\mathsf{E} = C_\zeta k^2$, the analogy between the TKE transport equation and Eq.~\eqref{eq:scalar} then becomes evident.

\section{Non-dimensionalization of the passive tracer transport equation}
\label{app:nondimensionalize}
In this work, the neural network is used to emulate the transport equation for the hypothetical volumetric concentration $\tau$ (dimensionless):
\begin{equation*}
    \bu \cdot \nabla \tau - \nabla \cdot (C_\nu \nabla \tau)  = C_g  \ell_{m}  \sqrt{\tau}  s^2 - C_\zeta \tau^2,
\end{equation*}
where $C_g$ (unit [\si{m^{-1} s}]), $C_\nu$ (unit [\si{m^2/s}]), and $C_\zeta$ (unit [\si{s^{-1}}]) are coefficients associated with production, diffusion, and dissipation, respectively. 
Although the PDE above is in a dimensional form with physical units, the inputs of the vector cloud neural network are all dimensionless quantities (see Table~\ref{tab:features}). Therefore, the network can be trained on data from  transport equations from vastly different length- and velocity-scales, as long as they are dynamically similar. Furthermore, the similarity argument also allows the network to predict convection-diffusion-reaction flows that have different scales from the training flows if they are dynamically similar. In the following, we show how dynamic similarity can be achieved for the convection-diffusion-reaction PDEs above.

We introduce the characteristic length $L_0$ and the characteristic flow velocity $U_0$ to normalize the variables as follows:
\begin{equation}
\mathbf{u}^{*}=\frac{\mathbf{u}}{U_0}, \quad \ell_{m}^{*}=\frac{\ell_{m}}{L_0}, \quad \nabla^{*}=L_0 \nabla,
\label{eq:nondim-scheme1}
\end{equation}
where the dimensionless variables are indicated in superscript $*$, and accordingly the strain rate magnitude and coefficients are normalized as follows:
\begin{equation}
    s^{*}=\frac{L_0}{U_0} s, \quad
    C_\nu^*=\frac{C_\nu}{U_0 L_0}, \quad C_g^*=U_0 C_g , \quad C_{\zeta}^{*}=\frac{L_0}{U_0} C_{\zeta}.
    \label{eq:nondim-scheme2}
\end{equation}
As such, we can rewrite the transport equation in the following dimensionless form:
\begin{equation}
\mathbf{u}^{*} \cdot \nabla^{*} \tau-\nabla^{*} \cdot\left(C_\nu^* \nabla^{*} \tau\right)=C_g^* \ell_{m}^{*} \sqrt{\tau} s^{* 2}- C_{\zeta}^{*} \tau^{2}.
\label{eq:nondim}
\end{equation}

From the non-dimensionlization scheme in Eq.~\eqref{eq:nondim-scheme2} and the resulting dimensionless equation~\eqref{eq:nondim}, we can see the solutions of two convection-diffusion-reaction equations at different scales are dynamically similar if they have the same non-dimensional diffusion, production, and dissipation coefficients ($C^*_\nu$, $C_g^*$, and $C^*_\zeta$) and the same dimensionless strain rate magnitude (i.e., the same time scale $T_0 \equiv U_0/L_0$). For example, the system studied in this work has length and velocity scales $L_0 = 1$~m and $U_0 = 1$~m/s. Consider system with a characteristic length $\lambda L_0$, it must have a characteristic velocity of $\lambda U_0$ and would further need to have the diffusivity $C_\nu$ scaled by a factor of $\lambda^2$, the production rate $C_g$ by $1/\lambda$, and the dissipation $C_\zeta$ by 1 in order to preserve dynamic similarity. 
Note that the characteristic length $L_0$ and the characteristic velocity $U_0$ must scale at the same ratio to achieve the dynamic similarity, because the production model $\mathsf{P}$ depends on the strain rate $\mathbf{s}$.

\section{Invariance of the proposed constitutive neural network}
\label{app:proof}
In this work, we are interested in predicting the scalar quantity $\tau(\bx_0)$ at the location $\bx_0$ from data collected at $n$ neighboring locations $\bx_i, i=1,\dots,n$. 
We introduce our methodology in two dimensional space in consistency with our computation examples. Extending it to three dimensional case is straightforward following the same procedure.
At each point $\bx_i$ in the collection we have velocity $\bu_i=\bu(\bx_i)$ and a number of other scalar features, denoted together by $\bw_i=\bw(\bx_i)\in \bbR^{l'}$. 
We denote the region-to-point mapping by $g$, i.e.,
\begin{equation}
\tau(\bx_0)=g\Big(\big\{\bx_i, \bu(\bx_i), \bw(\bx_i)\big\}_{i=1}^n\Big).
\end{equation}
Given the underlying physical knowledge, we want this mapping to possess three types of invariance: translational, rotational, and permutational invariances.
We first give a formal definition of these properties.
\begin{itemize}
\item{\bf{Translational invariance}.} If we shift the coordinate system with a constant vector $\by$, the mapping in the new system should be the same
\begin{equation}
    \tau(\bx_0+\by)=g\Big(\big\{\bx_i+\by, \bu(\bx_i+\by), \bw(\bx_i+\by)\big\}_{i=1}^n\Big).
\end{equation}
\item{\bf{Rotational invariance}.} If we rotate the coordinate system through an orthogonal matrix $\cO\in \bbR^{2\times2}$, the mapping in the new system should be the same
\begin{equation}
    \tau(\cO\bx_0)=g\Big(\big\{\cO\bx_i, \cO\bu(\cO\bx_i), \bw(\cO\bx_i)\big\}_{i=1}^n\Big).
\end{equation}
\item{\bf{Permutational invariance}.} The mapping should be independent of the indexing of collection points, i.e., if $\sigma$ denotes an arbitrary permutation of the set $\{1,2,\dots,n\}$, we have
\begin{equation}
    \tau(\bx_0)=g\Big(\big\{\bx_{\sigma(i)}, \bu(\bx_{\sigma(i)}), \bw(\bx_{\sigma(i)})\big\}_{i=1}^n\Big).
\end{equation}
\end{itemize}
In addition, the data we collected at the neighboring points can be interpreted as a discretized sampling of the continuous fields $\bu$ and $\bw$, and we are interested in the general case where $n$ varies in a certain range among different regions.
So we also want the mapping $g$ to be applicable to different sampling numbers $n$.

Granted the fitting ability of neural networks, the key to a achieve a general representation satisfying the above properties is an {\textit{embedding}} procedure that maps the original input to invariance preserving components. We draw inspiration from the following two observations.
\begin{itemize}
\item{\bf{Translation and rotation}.} If we define a relative coordinate matrix
\begin{equation}
\label{eq:rel_X}
    \cX = 
    \begin{bmatrix}
        (\bx_1-\bx_0)^\top \\
        (\bx_2-\bx_0)^\top \\
        \vdots\\
        (\bx_n-\bx_0)^\top 
    \end{bmatrix}\in \bbR^{n\times2}
\end{equation}
then the symmetric matrix $\cX\cX^\top \in \bbR^{n\times n}$ is an over-complete array of invariants with respect to translation and rotation, i.e., it contains the complete information of the pattern of neighboring points' locations~\cite{bartok2013representing,weyl1946classical}.
However, this symmetric matrix depends on the indexing of points. It switches rows and columns under a permutational operation of the neighboring points.

\item{\bf{Permutation and flexibility with sampling number}.} It is proven ~\cite{zaheer2017deep,han2019universal} that if $f$ is a function taking the set $\{\bz_i\}_{i=1}^n$ as input, i.e., $f$ is permutational invariant with respect to the set's elements, then it can be represented as
\begin{equation}
\label{eq:perm_decomposition}
    \Phi\Big(\frac1n\sum_{i=1}^n\phi(\bz_i)\Big)
\end{equation}
where $\phi(\bz_i)$ is a multidimensional function, and $\Phi(\cdot)$ is another general function. On the other hand, we can see such a representation intrinsically preserves the permutational invariance and flexibility with the number of point $n$.

This result is similar in flavor to the Kolmogorov-Arnold representation theorem~\cite{kolmogorov1957representation} but specific to permutational invariant functions. Here we give two examples to provide readers some intuition. ${z_1}^2+{z_2}^2+z_1z_2$ is permutational invariant with two scalar input variables $z_1,z_2$. This function can be written in the form of Eq.~\eqref{eq:perm_decomposition} when we define $\phi(z)=[z,z^2]^\top$ and $\Phi([a,b]^\top)=2a^2+b$:
\begin{equation}
    \Phi\Big(\frac12\sum_{i=1}^2\phi(\bz_i)\Big) = \Phi\bigg(\frac12\Big[\begin{smallmatrix}
   z_1+z_2 \\
   {z_1}^2+{z_2}^2
   \end{smallmatrix}\Big]\bigg) = 
   \frac{1}{2} {(z_1+z_2)}^2 + \frac{1}{2} ({z_1}^2+{z_2}^2) = 
   {z_1}^2+{z_2}^2+z_1z_2.
\end{equation}

Similarly, $z_1+z_2+z_3+2z_1z_2z_3$ is permutational invariant with three scalar input variables $z_1,z_2,z_3$. Assuming $\phi(z)=[z,z^2,z^3]^\top$ and $\Phi([a,b,c]^\top)=9a^3-9ab+3a+2c$, we can verify that the composition in the form of Eq.~\eqref{eq:perm_decomposition} gives us the desired function.
\end{itemize}

Inspired by the above observations, we construct our nonlocal mapping through neural networks in the following four steps. As explained below, the final output inherits all the invariant properties intrinsically.
\begin{enumerate}[(1)]
\item
    Given the input data $\big(\bx_0, \big\{\bx_i, \bu(\bx_i), \bw(\bx_i)\big\}_{i=1}^n\big)$, we define the input data matrix:
    \begin{equation}
    \cQ = [\cX, \cU, \cC]\in \bbR^{n\times(4+l')},    
    \end{equation}
    where
    \begin{equation}
        \cU = 
        \begin{bmatrix}
            \bu(\bx_1)^\top \\
            \bu(\bx_2)^\top \\
            \vdots \\
            \bu(\bx_n)^\top
        \end{bmatrix}\in \bbR^{n \times 2},\quad 
        \cW = 
        \begin{bmatrix}
            \bw(\bx_1)^\top \\
            \bw(\bx_2)^\top \\
            \vdots \\
            \bw(\bx_n)^\top
        \end{bmatrix} \in \bbR^{n \times l'}, \quad 
    \end{equation}
    and $\cX\in \bbR^{n\times 2}$ is defined in Eq.~\eqref{eq:rel_X}.
    Note that the elements in $\cW$ are all translational and rotational invariant.
\item
    We define $m$ embedding functions $\{\phi_k(\bw_i)\}_{k=1}^m$ for the scalar features $\bw_i$ of each point $i$ and compute the embedding matrix as
    \begin{equation}
    \cG = \begin{bmatrix}
         \phi_1(\bw_1) & \phi_2(\bw_1) & \ldots  & \phi_m(\bw_1)\\
          \phi_1(\bw_2) & \phi_2(\bw_2) &  \ldots & \phi_m(\bw_2) \\
          \vdots        & \vdots         & \ddots & \vdots \\
           \phi_1(\bw_n) & \phi_2(\bw_n) & \ldots & \phi_m(\bw_n) 
    \end{bmatrix} \in \bbR^{n \times m}.
    \end{equation}
    We can interpret matrix $\cG$ as a collection of basis, as seen in the last step below.
    %which is defined as a nonlinear function of the scalar quantity $\bw$.
    We also select the first $m'~(\leq m)$ columns of $\cG$ as another embedding matrix $\cG^\star \in \bbR^{n\times m'}$. Since all the features in $\bw$ are translational and rotational invariant, so are the elements in $\cG, \cG^\star$. In implementation we use an embedding network with $m$-dimensional output to instantiate the $m$ functions $\{\phi_k(\bw_i)\}_{k=1}^m$.
\item
    Next we consider
    \begin{equation}
    \cR\cR^\top= \cX\cX^\top + \cU\cU^\top + \cW\cW^\top \in \bbR^{n\times n},
    \end{equation}
    which is also translational and rotational invariant. In particular, to see its rotational invariance, we check that the new matrix under a rotation $\cO$ becomes
    \begin{equation}
    \cX\cO^\top\cO\cX^\top + \cU\cO^\top\cO\cU^\top + \cW\cW^\top,
    \end{equation}
    which remains the same.
    
\item
    Finally, we consider the encoded feature matrix
    \begin{equation}
        \cD = \frac{1}{n^2}\cG^\top\cR\cR^\top\cG^\star \in \bbR^{m\times m'}. \notag
    \end{equation}
    Given that all the elements in $\cG,\cG^\star, \cR\cR^\top$ are translational and rotational invariant, so are the elements of $\cD$. By the second observation above, we know each element in $\cG^\top \cR$ is permutational invariant, and so is $\cG^{\star \top} \cR$. So all the elements in $\cD$ possess all the desired properties. Now we reshape $\cD$ into a vector to form the input of the \emph{fitting} network $f_{\text{fit}}$, which outputs the predicted concentration $\hat{\tau}{(\bx_0)}$. 
\end{enumerate}
In the above procedures, essentially we need to train the parameters associated with two networks, the embedding network $f_{\text{embed}}$ and fitting network $f_{\text{fit}}$.
We also remark that by definition the evaluation of the predicted $\tau$ is flexible with the sampling number $n$ and has linear computational complexity with respect to $n$.

\section{A minimal example of invariance of the proposed constitutive neural network}
\label{app:mini-example}
In this example, we consider a minimal example where the scalar quantity $\tau$ is predicted based on the data collected at three neighboring locations and the input matrix $\cR$ has only three input features: 
\begin{equation}
\mathcal{Q} =\left[\begin{array}{lll}
x_{1} & y_{1} & c_{1} \\
x_{2} & y_{2} & c_{2} \\
x_{3} & y_{3} & c_{3}
\end{array}\right] \in \mathbb{R}^{3 \times 3} ,
\end{equation}
where $x$, $y$ are the relative coordinates and $c$ is the only scalar feature.
The scalar feature $c$ attached to each point is fed into the embedding network to obtain the embedding matrix $\cG$, which is processed individually and identically. For convenience, we set $m = 3$ and $m' = 1$. The embedding matrix and its submatrix are then computed as:
\begin{equation}
    \cG = \begin{bmatrix}
         \phi_1(c_1) & \phi_2(c_1)  & \phi_3(c_1)\\
          \phi_1(c_2) & \phi_2(c_2) & \phi_3(c_2) \\
           \phi_1(c_3) & \phi_2(c_3) & \phi_3(c_3) 
    \end{bmatrix} \in \bbR^{3 \times 3}, \quad
    \cG^{\star} = \begin{bmatrix}
         \phi_1(c_1) \\
          \phi_1(c_2) \\
           \phi_1(c_3)
    \end{bmatrix} \in \bbR^{3 \times 1}, \quad
\end{equation}
where $\{\phi_k(c)\}_{k=1}^3$ are embedding functions.
Finally, the invariant feature matrix is computed as:
\begin{equation}
\mathcal{D} = \frac{1}{3^2} \mathcal{G}^{\top} \mathcal{Q} \mathcal{Q}^{\top} \mathcal{G}^{\star} \in \mathbb{R}^{3 \times 1} ,
\end{equation}
where the feature matrix $\cD$ has translational, rotational and permutational invariances.

\begin{itemize}
\item{\bf{Translational invariance}.} 
Given that the input $x$, $y$ are the relative coordinates and $c$ is the passive scalar, the input matrix $\cQ$ will not change if the coordinate system is translated.

\item{\bf{Rotational invariance}.} 
Given a rotaion matrix $\cO$ defined as:
\begin{equation}
\cO=\left[\begin{array}{ll}
\cos{\theta} & \sin{\theta} \\
-\sin{\theta} & \cos{\theta}
\end{array}\right] \in \mathbb{R}^{2 \times 2} ,
\end{equation}
where $\theta$ is the rotation angle of the coordinate system. In the new rotated coordinate system, $\cG$ and $\cG^{\star}$ remain constant but the input matrix becomes
\begin{equation}
\cQ_{r} = 
\left[\begin{array}{lll}
x_{1} \cos{\theta}+y_{1} \sin{\theta} & -x_{1} \sin{\theta}+y_{1} \cos{\theta} & c_{1} \\
x_{2} \cos{\theta}+y_{2} \sin{\theta} & -x_{2} \sin{\theta}+y_{2} \cos{\theta} & c_{2} \\
x_{3} \cos{\theta}+y_{3} \sin{\theta} & -x_{3} \sin{\theta}+y_{3} \cos{\theta} & c_{3}
\end{array}\right] \in \bbR^{3 \times 3}.
\end{equation}
In this way, the pairwise projection is computed as:
\begin{equation}
\begin{split}
\cQ_{r} \cQ_{r}^{\top} &=
\left[\begin{array}{ccc}
c_{1}^{2}+x_{1}^{2}+y_{1}^{2} & c_{1} c_{2}+x_{1} x_{2}+y_{1} y_{2} & c_{1} c_{3}+x_{1} x_{3}+y_{1} y_{3} \\
c_{1} c_{2}+x_{1} x_{2}+y_{1} y_{2} & c_{2}^{2}+x_{2}^{2}+y_{2}^{2} & c_{2} c_{3}+x_{2} x_{3}+y_{2} y_{3} \\
c_{1} c_{3}+x_{1} x_{3}+y_{1} y_{3} & c_{2} c_{3}+x_{2} x_{3}+y_{2} y_{3} & c_{3}^{2}+x_{3}^{2}+y_{3}^{2}
\end{array}\right] \\
&= \cQ \cQ^{\top},
\end{split}
\end{equation}
which means that the pairwise projection stays the same after rotating the coordinate system. Therefore the feature matrix $\cD$ has rotational invariance.

\item{\bf{Permutational invariance}.} 
The obtained feature matrix $\cD$ should be independent of the indexing of three collected points. As defined above, $\mathcal{L} = \frac{1}{3} \mathcal{G}^{\top} \mathcal{Q}$, which is computed as:
\begin{equation}
\cL = \frac{1}{3}
\left[\begin{array}{lll}
\phi_{1}\left(c_{1}\right) x_{1}+\phi_{1}\left(c_{2}\right) x_{2}+\phi_{1}\left(c_{3}\right) x_{3} & \ldots & \phi_{1}\left(c_{1}\right) c_{1}+\phi_{1}\left(c_{2}\right) c_{2}+\phi_{1}\left(c_{3}\right) c_{3} \\
\phi_{2}\left(c_{1}\right) x_{1}+\phi_{2}\left(c_{2}\right) x_{2}+\phi_{2}\left(c_{3}\right) x_{3} & \ldots & \phi_{2}\left(c_{1}\right) c_{1}+\phi_{2}\left(c_{2}\right) c_{2}+\phi_{2}\left(c_{3}\right) c_{3} \\
\phi_{3}\left(c_{1}\right) x_{1}+\phi_{3}\left(c_{2}\right) x_{2}+\phi_{3}\left(c_{3}\right) x_{3} & \ldots & \phi_{3}\left(c_{1}\right) c_{1}+\phi_{3}\left(c_{2}\right) c_{2}+\phi_{3}\left(c_{3}\right) c_{3}
\end{array}\right] \in \bbR^{3 \times 3}.
\end{equation}
With regard to the first element $\cL_{11} = \frac{1}{3} \left[\phi_{1}\left(c_{1}\right) x_{1}+\phi_{1}\left(c_{2}\right) x_{2}+\phi_{1}\left(c_{3}\right) x_{3}\right] = \frac{1}{3} \sum_{i=1}^{3} \phi_{1}\left(c_{i}\right) x_i$, it is obviously independent of the indexing of collected points and so are other elements, which means the matrix $\cL$ is permutational invariant and so is $\cL^{\star}=\frac{1}{3} \mathcal{G}^{\star \top} \mathcal{Q}$. Therefore the feature matrix $\cD$ has permutational invariance.
\end{itemize}

\section{Network architectures and training parameters}
\label{app:nn}
Detailed architectures of the embedding neural network and the fitting network are presented in Table~\ref{tab:nn-details}. 
The Adam optimizer~\cite{kingma2015adam} is adopted to train the neural networks. The training process takes 2000 epochs with a batch size of 1024.
The training is scheduled such that the learning rate is initialized with 0.001 and is reduced by multiplying a factor of 0.7 every 600 epochs.

\begin{table}[htbp]
\caption{Detailed architectures of the embedding neural network and the fitting network.
The architecture denotes the numbers of neurons in each layer in sequence, from the input layer to the hidden layers (if any) and the output layer. The numbers of neurons in the input and output layers are highlighted in bold. Note that the embedding network operate identically on the scalar features $\bm{c}\in \bbR^{l'}$ associated with each of the point in the cloud and output a row of $m$ elements in matrix~$\cG$.
}
\centering
\begin{tabular}{P{4.2cm} P{3cm} P{4.5cm}}
\toprule[1pt]
 & Embedding network  & Fitting network ($\mathcal{D} \mapsto \tau$)\\
\midrule
No. of input neurons & $l' = 7 $ & $m\times m'= 256 \, (\cD \in \bbR^{64\times 4})$\\
No. of hidden layers & 2 & 1 \\
Architecture & 
(\textbf{7}, 32, 64, \textbf{64}) 
& 
(\textbf{256}, 128, \textbf{1})
\\
No. of output neurons & $m=64$ & 1 ($\tau \in \bbR$) \\
Activation functions & ReLU & ReLU, Linear (last layer) \\
No. of trainable parameters 
& 6528 & 33025
\\
\bottomrule[1pt]
\end{tabular}
\label{tab:nn-details}
\end{table}

% \clearpage

% \bibliography{nonlocal,hvsi,career,aiaa,reply-refer}

\end{document}